# Lewontin (1972)

Rasmus Grønfeldt Winther **0000-0001-8976-3052**

**Abstract**

Richard C. Lewontin is arguably the most influential evolutionary biologist of the second half of the 20th century. In this chapter, I provide two windows on his influential 1972 article "The Apportionment of Human Diversity": First, I show how the fourteen publications that he cites influenced him and framed his exploration; second, I present close readings of the five sections of the article: "Introduction," "The Genes," "The Samples," "The Measure of Diversity," and "The Results." I hope to illuminate the article's basic anatomy and argumentative arc, and why it became such a historically important document. In particular, I make explicit all of the mathematics (e.g., six Shannon information measures) and the general population genetic theory underlying this mathematics (e.g., the Wahlund effect). Lewontin did not make this explicit in his article. Furthermore, in redoing all of his calculations, I find that Lewontin made calculation errors (including rounding errors or omitting diversity component values) for all the genes he analyzed except one (P), and understated the among races diversity component, according to even just his own calculations. In reproducing the original computation, I find that the values of, respectively, within populations, among populations but within races, and among races diversity apportionments shift slightly (86%, 7%, 7%); here, in this "field guide" to Lewontin (1972), as well as in Winther (2022), I discuss this change in light of the values produced in subsequent replications of Lewontin's calculation with other statistics and data sets.

## Introduction

Richard C. Lewontin is arguably the most influential evolutionary biologist of the second half of the 20th century. A PhD student of Theodosius Dobzhansky—one of the architects of the neo-Darwinian modern synthesis along with R. A. Fisher, Sewall Wright, J. B. S. Haldane, Ernst Mayr, and George Gaylord Simpson—Lewontin has dazzled us with his experimental and mathematical prowess, conceptual sharpness, and inspirational qualities as a teacher, mentor, public speaker, and writer.

Of particular note for *Remapping Race in a Global Context* is his classic 1972 article, titled "The Apportionment of Human Diversity," especially the 85.4%/8.3%/6.3% distribution of genetic diversity components that Lewontin posits (1972, Table 4, p. 396) at three levels (within populations, among populations but within races, and among races). "Lewontin's distribution," as we could call it (see Winther, 2022), has been subject to wildly different ontological and political interpretations: Most commentators insist that it shows that we are all equal. However, some interlocutors claim that even relatively small percentages of average genetic difference among the aggregate, total population of respective continents—i.e., the so-called "races"—implies significant difference both in the evolutionary past (signature) and evolutionary future. This is not the place to discuss these matters, in part because they are covered elsewhere in this volume by A. W. F. Edwards, Lisa Gannett, Adam Hochman, Jonathan Michael Kaplan, Rasmus Nielsen, and Quayshawn Spencer, among others, as well as in the volume Postscript.

What I wish to do here is provide the reader two windows on Lewontin (1972). First, I summarize the publications that influenced him and framed his exploration. Lewontin collated an impressive array of knowledge on molecular and cytological genetics, taxonomy and natural history, statistical evolutionary theory, and conceptual (even philosophical) Darwinian insights. The bibliography of Lewontin (1972) sheds light on what concerned him in 1972.

The second window is a close reading of each of the five sections of the article: "Introduction," "The Genes," "The Samples," "The Measure of Diversity," and "The Results." This window illuminates the basic anatomy and argumentative arc of Lewontin (1972). In redoing all his calculations, I find that Lewontin's Distribution of (1) within populations, (2) among populations but within races, and (3) among races diversity apportionments are, respectively and rounding to the nearest percentage, in fact 86%/7%/7%. As I show, probably for the first time in a systematic manner, Lewontin made calculation errors (including rounding errors or omitting diversity values) for all the genes he analyzed except one. He also understated the among races diversity component, according to even just his own calculations.

Looking back 50 years later, we see that Lewontin (1972) crystallized a set of problems and questions that had been inchoate in the study of human evolution for a long time. It also provided partial answers, which have been immensely influential.

## Lewontin before Lewontin (1972)

The bibliography contains 14 publications in total, all of which are cited here.[1] In this section, I focus on eight of them that together present three interrelated themes providing critical context for our investigation. The publications are the five articles by Lewontin himself—three of them co-authored—together with articles by Theodosius Dobzhansky and Hermann J. Muller, as well as a technical report by Ladislav and Marie P. Dolanský. The themes are (1) the classical versus balance hypotheses of the extent and structure of genetic variation within species, (2) the molecularization of genetics, and (3) statistics and measures of genetic variation.

## Classical versus balance hypotheses

The importance of heritable variation for the evolutionary process cannot be overstated. But how is variation passed on, and how prevalent is it in different populations of a species? I will *not* here review some basic ingredients necessary to begin to answer this question: Gregor Mendel's principles of heredity, Thomas Hunt Morgan's confirmation of the chromosomal theory of inheritance, Francis Crick and James Watson's genes-are-DNA model, and the mathematical population genetics of Fisher, Wright, and Haldane, which delineates how evolutionary forces such as natural selection, genetic drift, migration, and mutation change gene frequencies in natural, experimental, and theoretical populations.

Lewontin (1972) represents a hallmark in investigating genetic variation in human populations. For the young Lewontin, the main question of experimental population genetics was: "At what proportion of his loci will the average individual in a population be heterozygous?" (Lewontin and Hubby, 1966, p. 603; cf. Hubby and Lewontin, 1966, p. 577). Understanding the depth and centrality of this question requires turning to the debate between classical and balance hypotheses, that is, between the idea that there is "a high level of heterozygosity in natural populations" and the view "that polymorphic loci will represent a small minority of all genes" (Lewontin and Hubby, 1966, p. 603).

The two key figures in this pugilistic mid-20th-century scientific controversy were Dobzhansky, Lewontin's mentor, and Hermann J. Muller, a fruit fly geneticist, student of T. H. Morgan and Nobel laureate for discovering that X-rays induce mutations. Lewontin (1972) cites Muller (1950) and Dobzhansky (1955). One of the earliest, simplest, and most definitive contrasts of the two hypotheses can be found in the latter article:

> According to the classical hypothesis, evolutionary changes consist in the main in gradual substitution and eventual fixation of the more favorable, in place of the less favorable, gene alleles and chromosomal structures. Superior alleles are established by natural selection, and supplant inferior ones. Most individuals in a Mendelian population should, then, be homozygous for most genes. Heterozygous loci will be a minority. …
>
> According to the balance hypothesis, the adaptive norm is an array of genotypes heterozygous for more or less numerous gene alleles, gene complexes, and chromosomal structures. Homozygotes for these genes and gene complexes occur in normal outbred populations only in a minority of individuals, and make these individuals more or less inferior to the norm in fitness.
>
> (Dobzhansky, 1955, p. 3)

Muller defended a stark view wherein natural selection was primarily directional and purifying, favoring one allele over all others at a given locus, and within a given environment. (Different populations and different "races" often live in distinct environments and niches.) In his 1950 article, Muller cites an earlier 1918 article of his in which he used "races" in a very generic sense synonymous with "varieties": "It is to the advantage of the organism that most genes shall be very stable, and present-day races are doubtless the products of a long process of selection" (Muller, 1918, p. 494, 1950, p. 122). That is, at every locus, selection eliminates all alleles except for the fittest one, thereby making most loci in a population homozygous.

Dobzhansky, in contrast, championed a more holistic picture in which most loci in most individuals were heterozygous—implying (due to Mendel's principles) that each locus requires many allele types, which can mix and match in different types of heterozygous pairs. For Dobzhansky, natural selection tended to be balancing selection, which favored heterozygotes over homozygotes (also sometimes, and perhaps confusingly, called "hybrid vigor").

Much rode on these two hypotheses, including distinct overarching pictures of the precise nature of selection, the evolutionary potential of natural populations, and the ways in which evolutionary forces such as selection, migration, and mutation interacted. This is not the place for a detailed discussion of classical versus balance hypotheses.[2] I simply wish to illustrate the larger context for why Lewontin cared to develop a population genetic research program focusing on assessing genic heterozygosity, which resulted in four articles with the telling, common, main title: "A Molecular Approach to the Study of Genic Heterozygosity in Natural Populations." Let us turn to the *molecular strategy* in order to understand Lewontin's overarching defense of his mentor's balance hypothesis.

## The molecularization of genetics

By the mid-60s, heritable variation had been assessed at the cellular level of chromosomal "inversions and translocations" (e.g., chromosomal segment rearrangements in *Drosophila*, cf. Winther, 2020, Figure 8.2, pp. 220–221), and even for "rare visible mutations at many loci" (Hubby and Lewontin, 1966, p. 577). However, what was required was a much more consistent, robust, and logical mode of pinpointing loci as well as allele variation at loci.

In their pathbreaking two articles from 1966, both cited in Lewontin (1972), Hubby and Lewontin devised a successful and influential molecular strategy for assessing both allele variation in a population and the typical amount of heterozygosity in individuals, within and among populations. They argued that such a strategy had to satisfy four logically iron-clad criteria:

> (1) Phenotypic differences caused by allelic substitution at *single loci* must be detectable in *single individuals*. (2) Allelic substitutions at one locus must be distinguishable from substitutions at other loci. (3) A substantial portion of (ideally, all) allelic substitutions must be distinguishable from each other. (4) Loci studied must be an unbiased sample of the genome with respect to physiological effects and degree of variation.
>
> (Hubby and Lewontin, 1966, p. 578)

Based on Hubby's earlier electrophoretic work (Hubby, 1963), Hubby and Lewontin articulate what I identify as a three-step molecular strategy. The first two steps were shown in Hubby and Lewontin (1966), whereas the third was an "application" (Hubby and Lewontin, 1966, p. 579), central especially to Lewontin and Hubby (1966):

1. Identify and distinguish distinct *Drosophila* proteins by different assaying procedures of purification (e.g., salting out, centrifuging), staining, and electrophoretic mobility, for each adult enzyme or larval protein.
2. Confirm the Mendelian inheritance of many of these proteins (and their associated protein variants), thereby identifying relevant alleles for each protein, and sometimes even the chromosome containing the locus coding for the protein. For instance, the locus for alkaline phosphatase-4, *ap*-4—now known as *Aph*-4—is sex-linked (Hubby and Lewontin, 1966, pp. 587–589).
3. Evaluate the allele frequencies of the protein alleles in five populations of *Drosophila pseudoobscura*: Flagstaff, Arizona; Mather, California; Wildrose, California; Cimarron, Colorado; Strawberry Canyon (Berkeley), California (Lewontin and Hubby, 1966, p. 596; most of these populations also provided the individual fruit flies for Hubby and Lewontin, 1966, p. 580). For instance, alkaline phosphatase-4 was found to have only two alleles—0.93 and 1.00, where the latter is the normalized distance the more frequent protein variant travels along the electrophoretic gel—with most populations fixed for allele 1.00 (Lewontin and Hubby, 1966, Table 1, p. 598, Table 2, p. 600).

Let us focus on the third step.

Lewontin and Hubby (1966) reported that 11 of the 18 loci they had identified and extensively studied across the five Arizona, California, and Colorado populations were monomorphic (p. 601). That is, 39% (i.e., 7/18) of loci in this study were polymorphic "over the whole species," "the average population is polymorphic for 30% of all loci," and the proportion of heterozygote loci in an individual *Drosophila*'s genome was, on average, 12% (p. 608). While these main results were perhaps somewhat less than expected under Dobzhansky's balance hypothesis, they were undoubtedly higher than predicted by Muller's classical hypothesis. Moreover, Lewontin and Hubby identified a series of "biases" in their own study, which were unavoidable at the time and caused them to systematically underestimate the proportions both of polymorphic loci and of heterozygote loci per individual (e.g., "electrophoretic separation detects only some of the differences between proteins," Lewontin and Hubby, 1966, p. 604). Interestingly, Lewontin (1967a) had used gene frequencies from 33 blood groups in the English population as reported in Race and Sanger (1962) (a resource he did not use in 1972), finding approximately 33% polymorphic loci and an average heterozygosity per individual of about 0.16, a "remarkabl[y] similar" result to Lewontin and Hubby (1966) (Lewontin, 1967a, p. 685). Indeed, Prakash et al. (1969) also found similar results. (Both Lewontin, 1967a, and Prakash et al., 1969, are cited in Lewontin, 1972.)[3]

Lewontin and Hubby (1966) posited five "categories" or kinds of loci: *monomorphism* (11 loci), *widespread polymorphism with one allele in high frequency* (three loci), *ubiquitous* *polymorphism with no wild type* (three loci), *local indigenous polymorphism* (one locus), and *local pure races* (0 loci). Note that "widespread" implies a lower prevalence than "ubiquitous," and "wild type" is a high-frequency and geographically pervasive allele. A "local pure race," which was nonexistent in their data set, would correspond to "populations homozygous for one allele and other populations homozygous for a different one" (Lewontin and Hubby, 1966, p. 602), which would have been very much in line with the classical hypothesis. The names and relative distributions of these categories resonate with the questions and problematics of Lewontin (1972).[4]

But we require one last framing theme, based on an obscure citation to a technical report out of MIT (Massachusetts Institute of Technology)'s Research Laboratory of Electronics. After all, Lewontin needed a *measure*, a *statistic*, for his analysis.

## Statistics and measures of genetic variation

Lewontin required a way to measure the total amount of genetic variation at different levels of population structure: within local populations, among populations but within continental races, and among races. Which measure of genetic variation (diversity) should we use, and how should we assess the relative amount of variation at the three different levels? How should we "apportion human diversity"?

Standard population genetic theory often starts with the measure of heterozygosity, $h$: the total number of heterozygotes of any kind, in a population, at a locus. Lewontin was well aware of this measure of genetic variation (1972, p. 388; eq. 1.1). (Incidentally, statisticians and ecologists often refer to this measure as *Gini* or *Gini diversity*, per Gini, 1912, cf. Simpson, 1949, work with which Lewontin possibly was familiar.) However, Lewontin preferred another one:

the Shannon information measure, *H*, which "bears a strong resemblance numerically to *h* [heterozygosity]." After all, this measure "is widely used to characterize species diversity in community ecology, and since I am performing a kind of taxonomic analysis here, I will use *H*" (Lewontin, 1972, p. 388). While it would be difficult to trace exactly the sources of Lewontin's choice of measure here, the work, and even the friendship, of ecologists such as Richard Levins (Lewontin's colleague at the University of Chicago before they both moved to Harvard University) and Robert MacArthur, among others, with their wealth of ecological knowledge and insight, would likely have influenced his decision.

In his calculations, Lewontin availed himself of a log table collated by two researchers, Ladislav Dolanský and Marie P. Dolanský, at least one of whom was associated with MIT's Research Laboratory of Electronics (RLE).[5] (The founder of information theory, Claude Shannon, had received his masters and PhD from MIT prior to WWII, and he returned to MIT as visiting professor in 1956, and, starting in 1957, became Professor of Communication Science and Professor of Mathematics at MIT; see Soni and Goodman, 2017, pp. 223–225.) The Shannon measure, also known as *Shannon entropy* or simply *entropy*, requires adding logarithms (in this case, in base 2)—that is, adding many $-p\log_2 p$—for each allele frequency, at each locus. As Lewontin stated, in an era just before sufficiently versatile pocket calculators, "the calculation of *H* is somewhat eased by published tables of [$-p\log_2 p$] (Dolanský and Dolanský, 1952)" (1972, p. 388).[6]

Here we get to the second part of the question with which I started this section. Lewontin had to calculate the Shannon measure, *H*, at three levels, as we shall see below. Calculating *H* was a steppingstone for apportioning human genetic diversity since, eventually, *H* at different levels had to be compared.

Lewontin's method of comparing *H* within populations, among populations but within races, and among races resembles Sewall Wright's *F*-statistics.[7] Even so, Lewontin's bibliography fails us in tracing this influence on Lewontin (1972). Admittedly, Dobzhansky (1955), contained in Lewontin's bibliography, cites relevant work by Wright on "panmictic units" or "demes" to articulate the notion of "Mendelian populations," which Dobzhansky took to exist also in *Homo sapiens*, grounded in "geographic isolation" and "marriage regulation" (1955, p. 2).

But more generally, and more importantly, Lewontin was familiar with Wright's rich and diverse work, including his mathematics of population structure and inbreeding coefficients, captured in his *F*-statistics. After all, Wright was a towering figure in population genetics and had also published with Lewontin's PhD mentor. Lewontin (1967b) described all of this, and more. Lewontin concluded a brief review of a 1950 book by Fisher as follows: "That Fisher could have completed a manuscript on the theory of inbreeding in 1961 without a single mention of Sewall Wright… bears witness to the power of pride and prejudice" (Lewontin, 1965, p. 1801). Wright's work attuned geneticists, including Lewontin, to the importance of population structure and to ways of measuring it.

Bibliographic chasing provides an excellent way to understand Lewontin (1972). The classical versus balance debate framed his understanding of what was at stake in detecting genetic variation in a variety of species, including *Homo sapiens*. And Lewontin was profoundly sensitive to the centrality of detecting genetic variation at the molecular level, as attested to by his work with Jack Hubby.

How do we use molecular genetic data to apportion diversity? We require a measure and a method of partitioning genetic variation. The Shannon measure—fed into an *F*-statistics-like apportionment of human diversity—was the last *mathematical* ingredient Lewontin deployed to answer what, at one point at least in Lewontin (1972), he took to be his "question": "How much of human diversity between populations is accounted for by more or less conventional racial classification?" (p. 386). The short answer? Not much, something like 6.3% [*sic*]. The longer answer? Let us revisit Lewontin's classic article using our three framing themes, addressing each section in turn.

## "Introduction"

Lewontin (1972) starts by discussing the "nodal" nature of variation. That is, "individuals fall in clusters in the space of phenotypic description." If we imagine an abstract phenotypic space—sometimes called a *morphospace*—with each dimension being a particular phenotypic character or aggregate set of characters, and each organism is a point in this space, then the organisms of a species are not evenly distributed throughout this space. Instead, Lewontin admits that there are three different levels of nodal or clustered variation in *Homo sapiens*: demes or populations, races, and our species as a whole. He also suggests that for many, but not all, "post-Darwinians," such multilevel nodal structure is a necessary "outcome of an evolutionary process" that takes genetic variation as its fuel (Lewontin, 1972, p. 381).

Lewontin (1972) can be interpreted as asking the following question: *To what extent are such nodes or groups, in particular human demes (populations) or human races, real*? Are such groups indicative of significant structuring of genetic variation, or are they imposed and reified, based on our perceptual biases and on classificatory expectations and norms in colonialist and racist societies?

In a nutshell, are races genetically real?

Our first two framing themes help us here. Start with the classical versus balance hypotheses. According to the classical view, "most men were homozygous for wild-type genes at virtually all their loci," such that "the obvious genetical differences in morphological and physiological characters between races are a major component of the total variation within the species" (Lewontin, 1972, p. 381). Muller held that different populations and different races (in organisms in general, not just humans) had been under a long process of selection, leaving most (if not all) loci fixed, that is, homozygous. Since the alleles fixed at particular loci differ among populations and among races, most (if not all) genetic variation is among populations or races, rather than within populations. In contrast, according to the balance view, attributed by Lewontin to Dobzhansky (1955), "heterozygosity is the rule in sexually reproducing species," such "that population and racial variations are likely to be less significant in the total species variation" (Lewontin, 1972, p. 382). The difference is immense—populations and races would be genetically real under the classical view, but according to the balance view, they would be mere cognitive or social (or both) epiphenomena or reifications.

To be precise, and especially for humans, the classical hypothesis of H. J. Muller was committed to something like the following three positions (Lewontin, 1972, pp. 381–382):

1. Within a species, there are real phenotypic nodes or classes or groups or clusters of organisms, at populational and racial levels.
2. Each phenotypic node is associated with a certain set of alleles at certain loci; these alleles tend to be fixed differently in different nodes.
3. Such fixed loci are representative of the clear majority of loci in the different nodes or groups since all loci have been subject to the strong hand of directional selection.

Interestingly, Lewontin accepted the bulk of the first two positions. Later in his article, he defends "the undoubted existence of such [racial] nodes in the taxonomic space," insisting that

> no one would confuse a Papuan aboriginal with any South American Indian, yet no one can give an objective criterion for where a dividing line should be drawn in the continuum from South American Indians through Polynesians, Micronesians, Melanesians, to Papuans.
>
> (Lewontin, 1972, p. 385)

Recall also that he acknowledged that there were "obvious genetical differences" for certain characters between races (Lewontin, 1972, p. 381). Indeed, Lewontin did not need to accept or insist on either the ephemerality or subjectivity of populational or racial phenotypic nodes [i.e., deny (1)], or the non-existence, or predictive or explanatory weakness, of simple, one-to-one gene-character (or genotype–phenotype) mappings [i.e., deny (2); although he did deny the absolute fixation of alleles], to make his overall point that especially genetic races are illusions.

Rather, Lewontin vehemently opposed the classical view regarding position (3)—that the loci underlying the nodal phenotypes were representative of the genome as a whole. Without the check of the "objective quantification" of human genetic variation (Lewontin, 1972, p. 382), the clear cognitive and social biases deploying "obvious and well differentiated stereotypes" (p. 385), and emphasizing intergroup (i.e., interpopulational, interracial) as opposed to intragroup variation, at both the phenotypic and genotypic levels, could run rampant. We tend to posit populations, and especially races, based on "those characters to which human perceptions are most finely tuned (nose, lip and eye shapes, skin color, hair form, and quantity), precisely because they are the characters that men ordinarily use to distinguish individuals" (p. 382). This bias belies the third classical hypothesis position that loci informative of populational and, especially racial, classification (i.e., loci with extreme allele frequency differences among populations, and among races, ideally fixed in different ways in different populations and races) were representative of all loci. Nevertheless, while Lewontin (1972) is a clear argument against the third position, Lewontin actually accepts very much here with respect to the classical hypothesis—more than we would expect and more than he accepted later in his career.

The objective quantification of genetic variation and diversity lies in our second framing theme, the molecularization of genetics. Lewontin referred both to the methods and the results of "protein electrophoresis" (reviewed above) and "immunological techniques," which permit the "direct" and "objective" assessment and evaluation of genetic diversity, per locus, among human individuals and groups at various levels of population structure. Citing a number of studies using "objective techniques" and "older information on the distribution of human blood group genes," Lewontin argues that it is possible to estimate the relative amount of intragroup (or intranode) versus intergroup (internode) genetic variation or diversity in humans and thus to provide a "firm quantitative basis" to claims about the genetic reality of human racial groups (Lewontin, 1972, pp. 382–383).[8] Let us now explore the genes Lewontin studied.

## "The genes"

Lewontin availed himself of the extensive data on the genetics of (1) blood groups and (2) serum proteins and red blood cell enzymes.

Lewontin cites one of his own articles where he had shown that of 33 blood group systems (see Table 1, Lewontin, 1967a, p. 682; he states "35 or so" in 1972), only 15 "segregat[e]" with at least one "alternative form" (allele) with higher than 1% frequency in any human population. Of these, only "9 systems have been characterized in enough populations to make them useful for our purposes" (Lewontin, 1972, p. 383).[9] Blood group antigens were detected by immunological techniques, their phenotypic frequencies in different populations surveyed, and their genotypic and allele frequencies inferred and calculated. Calculations followed assumptions about the genetic behavior of each blood group (e.g., gene and allele number, complete or incomplete dominance and codominance) and about Hardy–Weinberg equilibrium.

As his sources for blood group information, Lewontin cites Mourant (1954), Mourant et al. (1958), and Boyd (1950) (1972, p. 383). While he never states the specific tables or pages he used in these books, it is clear that he relied on their data tables with direct allele frequency information, and that he also sometimes inferred allele frequencies from phenotypic data, also given in tables (e.g., Blackfoot Indians for Kidd, Mourant, 1954, Table 39, p. 410,[10] and the entire Lewis data set). Scouring these references reveals that most allele frequency data were taken from Mourant (1954). By my count, data for six of the nine blood groups were exclusively from this source (Table 1.1). Titles of Mourant's Chapters 2–6 name all nine groups (e.g., Chapter 6: "The *P*, *Lutheran*, *Kell*, *Duffy*, *Kidd*, and Other Systems," p. v). However, ABO allele frequency data were almost certainly also taken from Mourant et al. (1958); it's unclear why else Lewontin would cite that source, which only contains extensive amounts of ABO data. It is difficult to determine which populations Lewontin sampled from the many available to him in very long ABO tables in the three references. One piece of evidence that he used Boyd (1950), at least for ABO, is that in Table 23, p. 223, Boyd gives ABO data for Shoshone in Wyoming, a group mentioned in Lewontin's Table 2 (my Table 1.2), but not otherwise presented, as far as I can tell, in any other relevant table for the other 16 genes. Boyd (1950) only offered useful allele frequency tables for ABO, MNS, and Rh (Table 1.1).[11]

The second set of loci or genes was identified via electrophoretic techniques. These genes are discussed second in "The Genes" section, but listed first in both Tables 1 (p. 384) and 3 (pp. 390–394) of Lewontin (1972), and I too shall follow this latter convention (though not for Figures 1.4 or 1.5, where genes are alphabetized). Although Hubby and Lewontin pioneered electrophoresis in *Drosophila*, it was especially Harris (1966, 1970) who helped develop it for humans. And while Lewontin (1972) cites Harris (1970), he mentions that he took data on the electrophoretically identified eight genes from Giblett (1969). Lewontin failed to give explicit criteria for how or why he chose these eight genes, but even a skim of Giblett (1969) reveals that Lewontin *went for the tables*: The genes of serum proteins and red blood cell enzymes that he surveyed are precisely the ones for which Giblett (1969) has useful allele frequency tables (Table 1.1). Entire chapters of Giblett (1969) are also dedicated to various features of each of these eight genes, (primarily) one gene per chapter.[12]

<COMP: Place Table 1.1 Here>

Is there bias in gene choice here? Each gene Lewontin chose satisfied two reasonable selection criteria: (1) the gene has more than one allele, and the different alleles are sufficiently frequent in at least some human populations, and (2) reasonable to extensive geographical sampling had already been done for that gene. Lewontin could perhaps have surveyed population genetic data on hemoglobin and its variants (e.g., sickle-cell; thalassemia), which would likely have increased the among race diversity component, but the genetics here are complex, and there were no relevant and useful allele data tables in Mourant (1954), Boyd (1950), or Giblett (1969).[13]

## "The samples"

In trying to answer whether races are genetically real—whether racial classification has any statistical, genetic, or scientific validity—Lewontin faced the colossal task of synthesizing a vast amount of data from many human populations. He employed a seven-race classification (Table 1.2), listing the populations he drew upon. It is not always easy to determine which exact populations he used, however; nor what he did when multiple allele frequencies were indicated for the same population; nor how he aggregated and averaged the allele frequencies of (sub-)populations. Before identifying just a few issues solely for the case of haptoglobin, I would like to tackle two dilemmas directly addressed in Lewontin (1972).

First, he aims for a classification that is "*a priori* representativ[e] of the range of human diversity" (Lewontin, 1972, p. 384). But how many populations should one sample, and how should one weight different populations with (immensely) different numbers of individuals? He ends up counting each population equally (see eqs. 1.7–1.9) and includes "as much as possible, equal numbers of African peoples, European nationalities, Oceanian populations, Asian peoples, and American Indian tribes" (p. 385). (How he names and refers to these populations is itself of interest, as is the fact that Australian aborigines are not here mentioned.) These choices lead to a bias toward overestimating both the total human genetic diversity and the among populations diversity component as opposed to the within populations component (see note 21).

The second methodological problem concerns which racial classification system to start from. Consider the following: Are Sámi (Lapps) Europeans or East Asians? How African are African Americans? More generally, should we (1) use "linguistic, historical, cultural, and morphological" "external" evidence, thus decreasing the among races diversity component by pouring and lumping genetically distinct populations into the same race? Or should we (2) use entirely genetic evidence to delineate populations and races, though this "has no end" as every population would have to be made a race? Lewontin opts for something quite close to (1), though he does make "a few switches based on obvious total genetic divergence" (Lewontin, 1972, p. 386).

<COMP: Place Table 1.2 Here>

Even a partial investigation of just one gene—haptoglobin—brings to light issues with Lewontin's analysis. The first cell of Table 3 of Lewontin (1972) states 25 as the number of European populations from which allele frequencies were gathered for haptoglobin. However, in counting populations in Giblett's (1969) Table 2.1, pp. 94–95, only 21 of his European populations match Lewontin's Table 2 (my Table 1.2). Lewontin does not list, for example, Sicilians, Sardinians, Yugoslavians, or Australians in Table 1.2. The allele frequencies of all Italian groups here are so close it is hard to tell how *data abstraction* (Winther, 2020, Chapter 3) was performed. Did he lump Sicilians and Sardinians with Italians or did he simply ignore them? Similarly, Lewontin asserts the use of 21 African populations (Table 3), but comparing Giblett's Table 2.1 to Table 1.2 gives only 19 African populations. Why are the Yoruba and Ibadan, which are listed under Nigerians in Giblett, 1969, p. 95, not listed in Table 1.2, whereas the other three Nigerian ethnicities listed by Giblett for haptoglobin are—Fulani, Habe, and Ibo? Are African Americans (p. 96) included under African here—and should they be? (I counted them as such, per Table 1.2.) Why are Ethiopians or the Kgalagadi from Giblett's table not listed in Lewontin's Table 2?

There are inordinately many small questions about how Lewontin abstracted from the data tables to his statistical tables.[14] After some checking of the mapping of allele frequencies from data sources to Lewontin's Table 3 across many of the genes, I have come to accept that Lewontin's data abstraction is roughly accurate, even if there are many outstanding questions and concerns.[15] What interests me more, and what I turn to in the last two sections, is the actual population genetic theory behind the apportionment of human diversity, and the recalculation of—and discovery of multiple errors in—Lewontin's Table 3.

## "The measure of diversity"

Let us explore how Lewontin actually performed his calculations. This section reviews the logic of his method, while the subsequent section examines Lewontin's results.

### A thought experiment

Simply in order to motivate intuitions about the three diversity components, consider three families of extreme and idealized cases:

1. Only within populations diversity: If all allele frequencies—and their associated diversities, whether *h* (heterozygosity) or *H* (Shannon measure)—are the same in every population of every race, then all heterozygosity or entropy is *within populations*. (Assuming, of course, that at least one locus has two or more alleles, neither of which is fixed.)
2. Only among populations but within races diversity: If, within a given continental race, each and every locus is fixed, for each population (that is, there is no within populations genetic diversity) such that not all populations of that race are fixed in the same way—that is, at least two populations are fixed for different alleles at at least one locus (and, ideally but not necessarily, any pairwise comparison of populations shows them fixed for different alleles at many loci),[16] *and* if that same, proportional distribution of differently fixed populations within a given race is repeated across all the races (i.e., no among races genetic variation), then all genetic diversity (heterozygosity or entropy) is *among populations, within races*.

3. Only among races diversity: If, for each race, every population within that race is identically fixed, across all loci (thereby eliminating heterozygosity or diversity, both within and among populations), *and* if the different races are not fixed in the same way—that is, at least two races are fixed for different alleles at at least one locus (and, ideally but not necessarily, any pairwise comparison of races shows them fixed for different alleles at many loci), then all heterozygosity or entropy would be *among races*.

### Two overarching measures of diversity

The only two explicit equations of the Lewontin (1972) "The Measure of Diversity" section are for the genetic diversity measures of heterozygosity ($h$) and Shannon information ($H$).

Lewontin formalizes the first as follows:

$$h = \sum_{i,\, j=1}^{u} p_i p_j, \quad (1.1)$$

where $p$ is allele frequency, in general; $p_i p_j$ is the genotype frequency of the $i$th and $j$th allele; and there are a total of $u$ alleles. Since allele types are rank-ordered (1, 2, 3, 4, …, $u$), there are two ways to get the same heterozygote: $p_1 p_2$ and $p_2 p_1$. (Visualize the ordinary, biallelic Punnett square in which both the mother and father are heterozygous.) And per standard Hardy–Weinberg calculations even for multiple alleles, the total genotype frequency of the $i$th and $j$th heterozygote is always twice the $p_i p_j$ genotype frequency. This can be inferred geometrically by comparing heterozygosity ($h$; Gini) in Figure 1.2 to $p(1-p)$ in Figure 1.1.

However, an alternative form of eq. 1.1 is[17]:

$$h = 1 - \sum_{i=1}^{u} p_i^2, \quad (1.2)$$

where $p_i^2$ is the homozygotic genotype frequency of the *i*th allele, and there are a total of *u* alleles. In general, for multiple alleles, eq. 1.2 is preferred to eq. 1.1, as it is easier for calculations and for the tracking of genotypes.

Regarding Shannon information, formally, Lewontin writes:

$$H = -\sum_{i=1}^{u} p_i \log_2 p_i, \qquad (1.3)$$

where $p_i$ is the frequency of the *i*th allele, the base of the (binary) logarithm is 2, and there are a total of *u* alleles. Above, it was mentioned that Lewontin preferred to use this entropy measure.

<COMP: Place Figure 1.1 Here>

<COMP: Place Figure 1.2 Here>

Diversity is most naturally thought of as a measure of a system's heterogeneity, potentially assessed at various hierarchical levels. As justification for the use of these two measures, Lewontin declaims four criteria that a diversity measure must meet (the criteria names are mine): (1) *minimum diversity criterion*: the measure should be at its minimal value, ideally 0, when there is only one allele in the population (or race or species), or when one allele of several is fixed; (2) *maximum diversity criterion*: the measure should be at its maximal value when all allele frequencies at a locus are equal, that is, when $p_i = p_j = p_k \ldots = p_u$, and the diversity should decrease as one or more alleles become rare (and others become common)[18]; (3) *allele number sensitivity criterion*: assuming, for simplicity's sake, equal allele frequencies, diversity increases as we increase the number of alleles: "a population with ten [equally frequent] alleles is obviously more diverse… than a population with two [equally frequent] alleles" (Lewontin, 1972, p. 388); and (4) *convexity criterion*: we will explore this criterion below. Both *h* and *H* satisfy these four conditions, as we can see in Figure 1.2, which represents a minimum diversity value at $p = 0$ or $p = 1$ and a maximum diversity value at $p = 0.5$, and is convex.

## Six diversity measures

Although he did not explicitly write out any of the following six diversity measures, Lewontin (1972) effectively calculated each of them for each gene of the 17 genes studied.[19]

$H_O$, the diversity of a given population:

$$H_O = -\sum_{i=1}^{u} p_{i,m} \log_2 p_{i,m}, \qquad (1.4)$$

where $p_{i,m}$ is the frequency of allele *i* in population *m*, and *u* is the total number of alleles at the particular locus.

$H_{race}$, the racially averaged diversity:

$$H_{race} = -\sum_{i=1}^{u} \bar{p}_{i,r} \log_2 \bar{p}_{i,r}, \qquad (1.5)$$

where $\bar{p}_{i,r}$ is the average frequency of the *i*th allele, within a given race, *r*, whereas *u* is the total number of alleles at the particular locus.[20] Lewontin actually calculates the (per allele, per locus) $\bar{p}_{i,r}$ by averaging the allele frequencies $p_{i,m}$ of each population of that race, effectively "counting each population once" (Lewontin, 1972, p. 389). This diversity value is calculated for each race independently.

$H_{species}$, the species-averaged diversity:

$$H_{species} = -\sum_{i=1}^{u} \bar{p}_{i,s} \log_2 \bar{p}_{i,s}, \qquad (1.6)$$

where $\overline{p}_{i,s}$ is the average frequency of the *i*th allele, within the entire species, again counting every single population once, and *u* is the total number of alleles at a locus. Lewontin actually calculates the (per allele, per locus) overarching $\overline{p}_{i,s}$ as a weighted average in which each race's $\overline{p}_{i,r}$ is weighted according to the number of populations it has $(N_r)$, out of the total species population number $(N_s)$, for that locus.

$H_{pop}$, the average population diversity of a race:

$$H_{pop} = \frac{1}{M}\sum_{m=1}^{M} H_{O,m}, \quad (1.7)$$

where $H_{O,m}$ is the population diversity $H_O$ of the *m*th population of a race, and *M* is the total number of populations in the given race. This diversity value is calculated for each race independently.

$\overline{H}_{pop}$, the weighted average of every race's $H_{pop}$:

$$\overline{H}_{pop} = \frac{1}{N_s}\sum_{r=1}^{R} N_r H_{pop,r}, \quad (1.8)$$

where $H_{pop,r}$ is $H_{pop}$ of the *r*th race, $N_r$ is the number of populations in race *r* (corresponding to *M* in eq. 1.7, which often differs for each race), $N_s$ is the total number of populations in the species, and *R* is the total number of races of the species.

$\overline{H}_{race}$, the weighted average of every race's $H_{race}$:

$$\overline{H}_{race} = \frac{1}{N_s}\sum_{r=1}^{R} N_r H_{race,r}, \quad (1.9)$$

where $H_{race,r}$ is $H_{race}$ of the *r*th race, $N_r$ is the number of populations in race *r*, $N_s$ is the total number of populations in the species, and *R* is the total number of races of the species.

From these six Shannon information diversity measures, the magic of Lewontin's analysis emerges. Here's how.

Eqs. 1.4–1.6 are explicitly Shannon information measures, using three different allele frequencies: $H_O$ uses allele frequency data of a single population, which is read from the tables reviewed in the section "The Genes" (this is the only measure of the six *not* listed in Table 3 of Lewontin, 1972); $H_{race}$ uses the original data table allele frequencies within populations, averaging them over populations, within a race (and is listed for each gene in Table 3 of Lewontin, 1972); and $H_{species}$ averages data table population allele frequencies over the entire species (and is listed for each gene in Table 3 of Lewontin, 1972). Furthermore, always counting each population once (i.e., equally), Lewontin calculated three kinds of averages (eqs. 1.7–1.9): $H_{pop}$, $\bar{H}_{pop}$, and $\bar{H}_{race}$.

Now, since eqs. 1.4–1.6 use different allele frequencies, three different diversity values will be produced. Because they are assessed and calculated at different levels, with more diversity or information (entropy) each time since we average allele frequencies at increasingly higher levels (viz., other populations within a race; other races), they are broadly independent of one another. There are metaphorically three degrees of freedom here—constrained by inequalities given below. This can be understood by observing that $\bar{H}_{pop}$ depends numerically directly on $H_{pop}$, which depends on $H_O$, whereas $\bar{H}_{race}$ is numerically rooted in $H_{race}$. Third, $H_{species}$ is not an average of explicit Shannon information diversity measures, but is a Shannon information diversity measure calculated directly from species-level allele frequencies, and is the highest diversity value of the six.[21]

**The Wahlund effect**

Importantly, $H_{race} \geq H_{pop}$, $H_{species} \geq \bar{H}_{race}$, and $\bar{H}_{race} \geq \bar{H}_{pop}$. We must understand why the inequalities hold in order to make sense of the apportionment of genetic diversity.

In population genetics, the *Wahlund effect*, discovered by the Swedish geneticist Sten Wahlund, states that there is often "excess homozygosity" of subdivided populations, relative to the total population. Specifically, the averaged heterozygosity (or $H_{pop}$) of two (or more) populations is almost always *lower* than the total heterozygosity (or $H_{race}$) calculated by *pooling* those populations into a grand total population and using the grand average allele frequencies across every population, at a given locus. To put it conversely and equivalently, the averaged homozygosity of subdivided populations is almost always *higher* than the average homozygosity of the total population. Indeed, the one exception (and the reason for the "almost always" in the previous two sentences) is that when all populations have the *same* allele frequencies for the relevant locus, then the heterozygosities (or entropies; cf. Figure 1.2) in the subdivided populations compared to the total, pooled population will be equal (i.e., not lower), but heterozygosities (or entropies) will never be higher in the subdivided populations.[22]

<COMP: Place Figure 1.3 Here>

The three inequalities above can be verified by turning to Figure 1.3. This figure represents the Wahlund effect geometrically, by comparing racially averaged diversity (eq. 1.5) to average population diversity of a race (eq. 1.7), for a biallelic locus. Figure 1.3 illustrates that $H_{race}$ will almost always be higher than $H_{pop}$, except when allele frequencies across populations are identical. In a nutshell: $H_{race} \geq H_{pop}$.[23] Differently put, the average allele frequency used to calculate $H_{race}$ (or, alternatively, $H_{species}$) will almost always give a higher diversity than taking the average of $H_{O1}$ and $H_{O2}$, that is, $H_{pop}$ (or, alternatively, $\bar{H}_{race}$).[24]

All of this is precisely because of the convexity criterion: "a collection of individuals made by pooling two populations ought always to be more diverse than the average of their separate diversities, unless the two populations are identical in *composition*" (Lewontin, 1972, 388). Observe that the convex diversity function $H_{race}$ (or, alternatively, $H_{species}$) almost always "overshoots" the line connecting the two populations, which must give the (averaged) $H_{pop}$ (or, alternatively, $\bar{H}_{race}$) exactly at the point on the line cutting allele frequency $\bar{p}_r$ of Figure 1.3, that is, $\bar{p}_{i,r}$ of eq. 1.5 (or, alternatively, $\bar{p}_{i,s}$ of eq. 1.6).[25] The only time this overshooting does not hold is when the populations have the same allele frequencies, and the vertical lines collapse together such that $H_{race} = H_{pop}$ (or, alternatively, $H_{species} = \bar{H}_{race}$).[26]

If all of this is true for the level depicted in Figure 1.3, it must *also* be true for measures averaging those levels, as captured, respectively, by eqs. 1.9 and 1.8. That is, $\bar{H}_{race} \geq \bar{H}_{pop}$.

## Apportioning diversity at three levels

With this framework in place, Lewontin apportions total, averaged diversity at three distinct levels. He does this with three explicitly stated equations (pp. 395–396), resonant with Wright's *F*-statistics[27]:

- Within populations diversity component:

$$\frac{\bar{H}_{pop}}{H_{species}}, \qquad (1.10)$$

- Among populations, within races:

$$\frac{\bar{H}_{race} - \bar{H}_{pop}}{H_{species}}, \qquad (1.11)$$

- Among races:

$$\frac{H_{species} - \bar{H}_{race}}{H_{species}}, \qquad (1.12)$$

And after calculating these diversity apportionments, which sum up to 1 (or 100%) per locus, Lewontin averages each of the three diversity components across all 17 loci (counting each locus equally, of course) to come up with the grand, averaged apportionment, the famous (but not quite accurate, even by his own Table 4) result of 85.4%, 8.3%, 6.3%.[28] To these results, I now turn.

## "The results"

### The general shape of global human genetic variation

Table 1.3 shows three genes—across three major continental regions—indicative of the full range of the 17 genes in terms of difference in allele frequencies among races. Lewontin's results imply that for common genes—that is, genes having at least two alleles, each with reasonably high allele frequencies in at least some populations (more than either 1% or 5%, per population genetic convention)—the gene tends to be globally distributed, and different human groups tend to have relatively similar allele frequencies.[29] Thus, the Duffy gene is an atypical example of a common gene, as it is more extremely diverged than average.[30] Similar to Rh, Duffy has an among race diversity component of approximately 26%—see eq. 1.12), Table 1.6, and Figures 1.4 and 1.5. In contrast, 6-Phosphogluconate dehydrogenase (6PGD) indicates less variation among populations than the average common gene. P is beautifully typical of common genes, showing some variation across continental regions.[31]

<COMP: Place Table 1.3 Here>

**A map key**

What do the different values indicated in Lewontin's (1972) Table 3, pp. 390–394, per gene, correspond to in the previous section's formalism?

The top row of *N* corresponds, for the indicated *N* of each race, to *M* (eq. 1.7) or to $N_r$ (eqs. 1.8 and 1.9), and, for the "Total" *N* of the species, to $N_s$ (eqs. 1.8 and 1.9). (I accepted Lewontin's assertions of *N*, and did not question them in my recalculations.)

The second row of $\bar{p}$ corresponds, for the indicated $\bar{p}$ of each race, to $\bar{p}_{i,r}$ (eq. 1.5), and for the first entry under the $H_{species}$ column,[32] to $\bar{p}_{i,s}$ (eq. 1.6). (I accepted Lewontin's assertions of $\bar{p}$, i.e., $\bar{p}_{i,r}$, for each race, and did not question them in my recalculations.)

The last three rows simply correspond, per race, to diversity values that are calculated, respectively, by eqs. 1.5 and 1.7 and by dividing the latter value by the former. (I accepted Lewontin's assertions of $H_{pop}$, i.e., eq. 1.7, per locus per race, and did not question them in my recalculations.)

Finally, in the last three columns of Table 3, the diversity values indicated are for $H_{species}$ (eq. 1.6), $\bar{H}_{race}$ (eq. 1.9), and $\bar{H}_{pop}$ (eq. 1.8).

Note that diversity values $H_O$ (eq. 1.4) are not given in Lewontin's table. They are only captured implicitly via eq. (1.7), in Lewontin's plain reporting of $H_{pop}$ values. This is one reason why it is so challenging to reverse-engineer the root populations *O* and their $H_{O,m}$ and $p_{i,m}$ values that Lewontin used for all his subsequent calculations.[33]

**Recalculating Lewontin**

Even if we set aside data abstraction issues already addressed, Lewontin commits calculation errors of various kinds in Table 3. These errors include miscalculated logarithms as well as inappropriate (weighted) averages of allele frequencies and diversity values. In other words, he does not always correctly implement eqs. 1.5, 1. 6, 1. 8, and 1. 9 (per the section immediately above, I could not check the implementation of eqs. 1.4 and 1.7). There are also some rounding errors, but these can be hard to determine and verify. I have checked the calculations for every gene multiple times in multiple ways, including manually, with all calculations done also in Excel. I did not use Dolanský and Dolanský (1952), though it would be interesting to do so. The errors I find in Lewontin's Table 3, together with the likely reason(s), are listed in Table 1.4. The interested reader is welcome to double-check my work, and contact me with any potential corrections. My Excel file is freely available online (Winther, 2021).

<COMP: Place Table 1.4 Here>

Table 1.5 is an example of how I recalculated Lewontin's Table 3, including always checking the diversity apportionments. The seven populations for Ag are: Swedes, Swiss, Finns, Italians, Thai, Japanese, and Indians (Giblett, 1969, Table 5.1, p. 181; cf. Table 1.1).

<COMP: Place Table 1.5 Here>

It is worth pointing out that for every gene with Table 3 calculation errors (except one), as listed in Table 1.4, the final diversity apportionment was also impacted.[34] I have therefore produced a revised table, in the same format as Lewontin's Table 4, "Proportion of Genetic Diversity Accounted for Within and Between Populations and Races" (1972, p. 396), with these recalculated results (Table 1.6). Importantly, while there are some rounding errors in Lewontin's calculations of Table 4 diversity apportionments from his Table 3, I do not track all of these here. That is, I do not here provide a revised Lewontin Table 4 based on his actual, though often incorrect, $H_{species}$, $\bar{H}_{race}$, and $\bar{H}_{pop}$ Table 3 values. After all, it is Table 4 on page 396 of Lewontin (1972) that became influential. Now, of the five genes with no calculation errors (including rounding errors),[35] as listed in Table 1.4, all except P are changed in Tables 1.6 and 1.7, as compared to Lewontin's Table 4. (Added errors between the tables can be identified because Excel keeps 15 significant digits of precision, more than sufficient for our purposes[36]; thus, sometimes, even if there are no calculation errors for a given gene in Lewontin's Table 3, the apportionment can still shift a slight amount up or down. In the cases of Kell and Lutheran, Lewontin made rounding errors in moving from his Table 3 to his Table 4; only in the case of P, did Lewontin not commit any errors of any kind.)

<COMP: Place Table 1.6 Here>

In a nutshell, Lewontin is effectively using each gene as an independent statistical test of whether race is real—that is, of whether racial classification is either predictive or explanatory of genetic differences among groups (Table 1.6). Lewontin wants to say that it is neither predictive nor explanatory.

<COMP: Place Table 1.7 Here>

For almost all of the 17 genes, except for P, Lewontin made at least one calculation error (including rounding errors). Only a few of these errors were serious, and they were not systematic (Figures 1.4 and 1.5). However, the overstatement (by 0.7%) of the among populations diversity component and the understatement (by 1.3%) of the among races components, relative to just his own Table 4, merit further discussion (Table 1.6). Since the recalculated among populations but within races and among races diversity components are almost the same—7.1/7.2%—I suggest that we rethink Lewontin's distribution as properly **86%/7%/7%** (Table 1.6). While this result may seem anticlimactic, it may be a better mnemonic. I, for one, am grateful for having been able to redo all his calculations in order to ensure that we are getting the calculations right in this area of inquiry, and in the interest of scientific reproducibility.[37]

<COMP: Place Figure 1.4 Here>

<COMP: Place Figure 1.5 Here>

## Conclusion

Lewontin (1972) deserves to be celebrated as it turns 50. In beginning work on what was intended to be a short introduction to the piece, its deep multidimensionality became quickly clear to me: evolutionary theory, statistics, politics, and ethics intertwine. As I started finding errors in his results, and as I began digging into his concise but illuminating bibliography, I realized that it would not be a simple matter to introduce this seminal work. I felt I had to describe Lewontin before Lewontin (1972), present the content—explicit as well as implicit—of the different sections of his article, and carefully check all his calculations and claims. The resulting chapter is a kind of field guide to Lewontin (1972).

This chapter does not attempt to cover everything. For instance, Lewontin, a Marxist, mentions on the first page of his article the importance of "long-term changes in socioeconomic relations" in shaping professional views on "the relative importance and extent of intragroup as opposed to intergroup variation" (Lewontin, 1972, p. 381, citing Lewontin, 1968), and he also believes he has shown "racial classification" to be "of virtually no genetic or taxonomic significance" (Lewontin, 1972, p. 397). More ironically, he accepts race at a phenotypic level, as we saw above; he is not as judgmentally harsh about the genetic reality of populations qua nodes or groups as he is about the genetic reality of races qua nodes or groups, although he finds that their diversity apportionments are commensurable, and even overstates the former diversity component at the expense of the latter, and two years later publishes a book (Lewontin, 1974) arguing for the entire genome as a unit of selection—even though in 1972 he unavoidably averages across loci, thereby losing classificatory information, as so many of the chapters in this volume, and elsewhere, discuss (e.g., Winther, 2018).[38]

In addition, there are potential ethical concerns in data collection and management, and in theoretical calculation, abstraction, and modeling. Almost certainly, there are any number of such concerns involved in the data collection and management of the thousands of research papers consulted by Mourant (1954) (e.g., 1716 numbered references, pp. 239–335[39]), and the many hundreds referred to by Giblett (1969), in mapping out their rich data tables of allele frequencies. These important ethical issues have been covered in this volume by Guillermo Delgado-P, Kelly Happe, and Krystal Tsosie. The extent and exact nature of Lewontin's culpability in such matters—indirect as it may be—is interesting and should be further investigated.

I hope I will be forgiven for not discussing these matters here, as they have been discussed extensively and perhaps exhaustively elsewhere. The particular responsibility of my chapter is to present, for the first time it would seem, the full flesh and bones of the scientific background, data, theory, and results of Lewontin (1972). After all, this classic is remarkably telegraphic. It neither explicitly lists data sources nor explains background population genetic theory. It is also replete with calculation errors—recall that Lewontin made errors for all genes except one.

Could I derive Lewontin's results from his data? Could I replicate and reproduce his analyses? In the end, his diversity component distribution among the three levels should be revised to **86%/7%/7%**. This is a small correction. While Lewontin's calculation errors are not systematic (Figures 1.4 and 1.5), this could not have been known before the recalculations presented in this chapter, nor could the fact that he overstates the among populations diversity component while understating the among races diversity component, even by his own Table 4 (Table 1.6). We should also recall that his results have (only) roughly held up to the test of time.[40] Fifty years on from Lewontin (1972), it is important to remember that, in addition to adulating a great work such as this one, we must also strive to understand and evaluate, on a deep level, its content.

## Acknowledgments

Figure 1.1 Entropy (*H*) and Gini (*h*) summand values. The values of the components (summands) of our two measures, plotted against allele frequency *p*, for a biallelic locus. (Concept and draft by Rasmus Grønfeldt Winther; illustrated by Amir Najmi using ggplot2 in R.)

Figure 1.2 Entropy versus Gini diversity. The actual diversity values of Shannon information, *H*, and heterozygosity, *h*, mapped against allele frequency, for a biallelic locus, per, respectively, eqs (1.3) and (1.1) or (1.2). (Concept and draft by Rasmus Grønfeldt Winther; illustrated by Amir Najmi using ggplot2 in R.)

Figure 1.3 The Wahlund effect per convexity criterion. Two populations are shown, and diversity is entropy, that is, the Shannon information measure. $H_{O1}$ and $H_{O2}$ are the diversities of the first and the second population, respectively. Since the locus is biallelic—i.e., the other allele frequency is simply $(1-p)$—and to avoid confusion both about which population's allele frequency is being depicted and about the fact that the frequency for the same allele is represented in both populations, $p_{i,m}$ is written as $p_{O1}$ and $p_{O2}$ for, respectively, the first and the second population. See text for further archaeology of the figure. (Concept and draft by Rasmus Grønfeldt Winther; illustrated by Amir Najmi using ggplot2 in R.)

Figure 1.4 Lewontin–Winther bar plots. Bar plots of the three diversity components for each gene (on the same scale), as presented in Lewontin's Table 4, p. 396, compared to the true recalculated values from Table 1.6. Differences are often minimal. (Concept and illustration by Amir Najmi using ggplot2 in R.)

Figure 1.5 Lewontin–Winther scatter plots. Scatter plots of the differences between Lewontin's Table 4 values and the correct values from Table 1.6. The diagonal line maps the identity of value between the two. Most deviations from identity are small (but note Ak). Scales differ among the three plots because of the different magnitude ranges of the three diversity components (although among populations and among races are commensurable in magnitude). The "Winther" component values for each gene (that is, the *x*-axis) can be checked against Table 1.6. Note that while the "within_pops" scatter plot shows 17 genes, the "among_pops" and "among_races" scatter plots show only 14 genes because Lewontin had not calculated those respective values for Ag, Lp, and Xm genes. (Concept and illustration by Amir Najmi using ggplot2 in R; this image appears as gray scale in the print book, in color in the eBook, and as a downloadable eResource from www.routledge.com/9781138631434 [Hardback ISBN: 9781138631434].)

**Table 1.1** Lewontin's 17 Genes

| **Locus/Gene** | **Data Source** |
|---|---|
| Serum Proteins and Red Blood Cell Enzymes | |
| Haptoglobin (Hp) | Giblett (1969), Table 2.1, pp. 94–98 |
| Lipoprotein Ag (Ag) | Giblett (1969), Table 5.1, p. 181 |
| Lipoprotein Lp (Lp) | Giblett (1969), Table 5.2, p. 184 |
| □$_2$ Macroglobulin (Xm) | Giblett (1969), Table 8.2, p. 257 |
| Red cell acid phosphatase (APh) | Giblett (1969), Table 11.1, pp. 436–437[1] |

[1] There are strictly speaking five alleles here, with the two least frequent ones significantly present only in African populations. Lewontin seems to have grouped these two as a single allele.

| | |
|---|---|
| 6-Phosphogluconate dehydrogenase (6PGD) | Giblett (1969), Table 13.2, pp. 492–493 |
| Phosphoglucomutase (PGM) | Giblett (1969), Table 14.1, pp. 506–507 |
| Adenylate kinase (Ak) | Giblett (1969), Table 15.1, pp. 516–517 |
| Blood Groups | |
| Kidd (Jk) | Mourant (1954), Table 39, p. 410 |
| Duffy (Fy) | Mourant (1954), Table 38, pp. 408–409 |
| Lewis (Le) | Mourant (1954), Table 36, p. 406; perhaps Table 37, p. 407[2] |
| Kell (K) | Mourant (1954), Table 34, pp. 402–404; perhaps Table 35, p. 405[3] |
| Lutheran (Lu) | Mourant (1954), Table 33, pp. 400–401 |
| P | Mourant (1954), Table 19, pp. 366–369 |
| MNS | Boyd (1950), Table 25, pp. 234–235; Mourant (1954), Table 17, pp. 358–364[4] |

[2] Population numbers and types match Table 36 more than Table 37. But it is evident from Lewontin (1972), Table 3, that he chose to ignore Indigenous Americans (Table 36) and African (Table 37) populations. Table 37 represents testing with two antisera rather than one, as in Table 36. Here I suspect Lewontin used Table 36 primarily, as allele frequencies are much easier to infer from it than from Table 37.

[3] Population numbers reasonably match Table 34. But it is possible that for supplemental data, Lewontin drew on the much smaller Table 35, which tested with two antisera rather than one, as Table 34 did.

[4] Mourant (1954), Table 18, p. 365, concerns MNS information only for Chippewas of Minnesota; Table 16 only presents MN information.

| | |
|---|---|
| Rh | Boyd (1950), Table 29, pp. 244–245; Mourant (1954), Table 21, pp. 377–381[5] |
| ABO | Boyd (1950), Table 23, pp. 223–225, Table 27, pp. 238–239, Table 31, p. 265; Mourant et al. (1958); Mourant (1954), Table 14, pp. 339–343[6] |

The genes or loci used in Lewontin (1972), together with the source of the allele frequency data tables.

**Table 1.2** "Inclusive List of All Populations Used for Any Gene in This Study by the Racial Classification Used in this Study"

| |
|---|
| **Europeans**<br>Arabs, Armenians, Austrians, Basques, Belgians, Bulgarians, Czechs, Danes, Dutch, Egyptians, English, Estonians, Finns, French, Georgians, Germans, Greeks, Gypsies, Hungarians, Icelanders, Indians (Hindi speaking), Italians, Irani, Norwegians, Oriental Jews, Pakistani (Urdu-speakers), Poles, Portuguese, Russians, Spaniards, Swedes, Swiss, Syrians, Tristan da Cunhans, Welsh |
| **Africans**<br>Abyssinians (Amharas), Bantu, Barundi, Batutsi, Bushmen, Congolese, Ewe, Fulani, Gambians, Ghanaians, Habe, Hottentot, Hututu, Ibo, Iraqi, Kenyans, Kikuyu, Liberians, Luo, Madagascans, |

[5] Mourant's Table 20, pp. 370-376, represents using only one antiserum and only has two alleles. Table 23, pp. 383-388, reflects using four antisera, and has eight genetic types. Three antisera were used in the studies collated for Table 21, which best represents the six alleles Lewontin has in his Table 3, although he admits that for Rh, the six allelic types or classes involve conflating subtypes or subclasses (1972, p. 383). Moreover, comparing some racially averaged allele frequencies from Table 3 verifies that his table nicely matches Table 21. (Table 22, p. 382, concerns South African "Bushmen" only. Mourant's Tables 24–32 also capture Rh information, but as the reader can verify, these tables become increasingly specialized in populations or in the number of genetic types, which increases significantly, or both. It is unlikely that Lewontin used them.)

[6] Lewontin may have used the much smaller Table 15, p. 344, of Mourant (1954) for supplemental data.

| |
|---|
| Mozambiquans, Msutu, Nigerians, Pygmies, Senegalese, Shona, Somalis, Sudanese, Tanganyikans, Tutsi, Ugandans, US Blacks, "West Africans," Xhosa, Zulu |
| **East Asians** |
| Ainu, Bhutanese, Bogobos, Bruneians, Buriats, Chinese, Dyaks, Filipinos, Ghashgai, Indonesians, Japanese, Javanese, Kirghiz, Koreans, Lapps, Malayans, Senoy, Siamese, Taiwanese, Tatars, Thais, Turks |
| **South Asian Aborigines** |
| Andamanese, Badagas, Chenchu, Irula, Marathas, Nairs, Oraons, Onge, Tamils, Todas |
| **Indigenous Americans** |
| Alacaluf, Aleuts, Apache, Atacameños, "Athabascans," Aymara, Bororo, Blackfeet, Bloods, "Brazilian Indians," Chippewa, Caingang, Choco, Coushatta, Cuna, Diegueños, Eskimo, Flathead, Huasteco, Huichol, Ica, Kwakiutl, Labradors, Lacandon, Mapuche, Maya, "Mexican Indians," Navaho, Nez Percé, Páez, Pehuenches, Pueblo, Quechua, Seminole, Shoshone, Toba, Utes, "Venezuelan Indians," Xavante, Yanomama |
| **Oceanians** |
| Admiralty Islanders, Caroline Islanders, Easter Islanders, Ellice Islanders, Fijians, Gilbertese, Guamanians, Hawaiians, Kapingas, Maori, Marshallese, Melanauans, "Melanesians," "Micronesians," New Britons, New Caledonians, New Hebrideans, Palauans, Papuans, "Polynesians," Saipanese, Samoans, Solomon Islanders, Tongans, Trukese, Yapese |
| **Australian Aborigines** |

Source: Lewontin (1972), Table 2, p. 387; I have silently fixed a number of infelicities in Lewontin's list.

**Table 1.3** 6PGD, Duffy, and P Allele Frequencies across Three Continental Regions or Races

| Gene | Europeans | Africans | East Asians |
|---|---|---|---|
| 6PGD | 0.961 | 0.914 | 0.905 |
| Duffy | 0.410 | 0.072 | 0.784 |
| P | 0.533 | 0.693 | 0.433 |

Source: Lewontin (1972), Table 3, pp. 390–394.

**Table 1.4** Lewontin (1972) Table 3 Calculation Errors

| **Gene/Locus** | **Lewontin (1972) Table 3 Errors?** |
|---|---|
| Serum Proteins and Red Blood Cell Enzymes | |
| Haptoglobin (Hp) | • Incorrect $\overline{p}_s$ (0.457 not 0.456) and $H_{species}$ (0.995 not 0.994), due to rounding errors |
| Lipoprotein Ag (Ag) | • None, although the indicated value for $\overline{H}_{race}$ is really for $\overline{H}_{pop}$<br>• I used the relevant data table, recalculated, and made diversity apportionments explicit |
| Lipoprotein Lp (Lp) | • Incorrect averaging in calculating $\overline{p}_s$ (0.159 not 0.162)<br>• Incorrect $H_{species}$ (0.632 not 0.639) due to former error, and incorrect $\overline{H}_{pop}$ (0.596 not 0.600) due to averaging error<br>• I used the relevant data table, recalculated, and made diversity apportionments explicit |
| □$_2$ Macroglobulin (Xm) | • None |
| | • I used the relevant data table, recalculated, and made diversity apportionments explicit |
| Red Cell Acid Phosphatase (APh) | • Incorrect $\bar{H}_{race}$ (0.976 not 0.977) and $\bar{H}_{pop}$ (0.918 not 0.917), due to rounding errors<br>• Incorrect $\bar{p}_{2,s}$ (0.682 not 0.683) and $\bar{p}_{4,s}$ (0.002 not 0.001), both likely rounding errors<br>• Incorrect $H_{species}$ (0.999 not 0.989) due to logarithm or rounding errors, or both[7] |
| 6-Phosphogluconate dehydrogenase (6PGD) | • Incorrect $\bar{H}_{pop}$ (0.287 not 0.286) due to a rounding error |
| Phosphoglucomutase (PGM) | • Incorrect $H_{species}$ (0.759 not 0.758) due to a rounding error |

[7] By "logarithm error," I mean any error Lewontin might have made in reading or calculating from the values listed in Dolanský and Dolanský (1952). They list the values of *p* for calculating binary or base two logarithms to three decimal figures, and the actual logarithmic values to six decimal figures. Interestingly, and likely because he followed this logarithm table, Lewontin presented all allele frequencies to the nearest thousandth. (Every diversity measure and subsequent diversity apportionment was also rounded to the nearest thousandth by Lewontin.) Depending on the gene, Gibblett (1969) presented allele frequency data to the nearest hundredth or thousandth, and Mourant (1954) generally to the nearest ten thousandth.

| | |
|---|---|
| Adenylate kinase (Ak) | • Incorrect logarithm in calculating $H_{race}$ for East Asians (0.118 not 0.095)<br>• Incorrect $\bar{H}_{race}$ (0.164 not 0.160) and $\bar{H}_{pop}$ (0.136 not 0.156), the former due to the logarithm error, the latter due to averaging error |
| Blood Groups | |
| Kidd (Jk) | • Incorrect averaging in calculating $\bar{p}_s$ (0.632 not 0.411)<br>• Incorrect $H_{species}$ (0.949 not 0.977) due to former error |
| Duffy (Fy) | • Incorrect $\bar{p}_s$ (0.646 not 0.645) due to a rounding error |
| Lewis (Le) | • Incorrect $H_{species}$ (0.995 not 0.994) due to a rounding error |
| Kell (K) | • None |
| Lutheran (Lu) | • None |
| P | • None |
| MNS | • The four allele frequencies of South Asian Aborigines add up to 1.03, not 1 (I did not change allele frequencies)<br>• Incorrect $H_{race}$ for Africans (1.694 not 1.695), South Asian Aborigines (1.784 |
| | not 1.785), and Indigenous Americans (1.534 not 1.609), due to logarithm or rounding errors, or both<br>• Incorrect $\bar{p}_{2,s}$ (0.421 not 0.420) and $\bar{p}_{4,s}$ (0.354 not 0.353), both likely rounding errors<br>• Incorrect $\bar{H}_{race}$ (1.655 not 1.663) and $\bar{H}_{pop}$ (1.581 not 1.591), the former due to prior $H_{race}$ errors, the latter due to averaging error |
| Rh | • The six allele frequencies add up to 1 only for Oceanians and Australian Aborigines (I did not change allele frequencies)<br>• Incorrect averaging in calculating $\bar{p}_{1,s}$ (0.517 not 0.518), $\bar{p}_{3,s}$ (0.147 not 0.148), and $\bar{p}_{5,s}$ (0.165 not 0.166), likely rounding errors<br>• Incorrect $H_{race}$ for Europeans (1.765 not 1.763), Indigenous Americans (1.51 not 1.509), and Australian Aborigines (1.601 not 1.600), due to logarithm or rounding errors, or both |
| | • Incorrect $\bar{H}_{pop}$ (1.293 not 1.281) due to an averaging error<br>• Incorrect $\bar{H}_{race}$ (1.421 not 1.420), due likely to both rounding and logarithm errors |
| ABO | • The sum of three allele frequencies for East Asians is 0.999 (I did not change allele frequencies)<br>• Incorrect $H_{race}$ for Africans (0.153 not 0.154) due to a logarithm or rounding error, or both<br>• Incorrect $\bar{H}_{pop}$ (1.145 not 1.126) due to an averaging error<br>• Incorrect $\bar{H}_{species}$ (1.240 not 1.241) likely due to a rounding error |

I had to recalculate Ag, Lp, and Xm from scratch, using the relevant data tables (from Giblett, 1969; cf. Table 1.1), to make explicit the diversity apportionment at each of the three levels, as Lewontin only listed the within populations diversity component for them (eq. 1.10). (That allele frequencies do not always add up exactly to one can be unavoidable due to prior, legitimate rounding of the frequencies; thus, I left these for MNS, Rh, and ABO, although for Rh there was likely more error than just this rounding error—e.g., for Indigenous Americans the six allele frequencies add up only to 0.989.

**Table 1.5** Author Excel Sheet of Ag Recalculations

| **Lipoprotein Ag** (Giblett, p. 181) | | | | |
|---|---|---|---|---|
| | p (EU-ans) | Ho (EU-ans) | p (EA-ans) | Ho (EA-ans) |
| | 0.23 | 0.7780113 | 0.69 | 0.89317346 |
| | 0.24 | 0.79504028 | 0.73 | 0.84146464 |
| | 0.31 | 0.89317346 | 0.74 | 0.82674637 |
| | 0.23 | 0.7780113 | | |
| N | 4 | | 3 | |
| Hpop | | 0.81105909 | | 0.85379482 |
| **Hpop-bar** | | 0.8293744 | | |
| p-bar (race) | 0.2525 | | 0.720 | |
| H race | 0.81521654 | | 0.85545081 | |
| **Hrace-bar** | 0.8324598 | | | |
| p-bar (species) | 0.45285714 | | | |
| **Hspecies** | 0.99357783 | | | |
| w/in pops | | 0.83473521 | | |
| Among pops/within race | | 0.00310534 | | |
| Among races | | 0.16215945 | | |

Source of data: Giblett (1969), 5.1, p. 181; cf. Table 1.1 above.

"EU-ans": European populations; "EA-ans": East Asian populations. Equations are embedded in the cells.

**Table 1.6** The True Genetic Diversity Apportionment

| | Within Pops | Among Pops | Among Races |
|---|---|---|---|
| Hp | 0.893 | 0.050 | 0.057 |
| Ag | 0.835 | 0.003 | 0.162 |
| Lp | 0.942 | 0.025 | 0.033 |
| Xm | 0.997 | 0 | 0.003 |
| APh | 0.919 | 0.059 | 0.023 |
| 6PGD | 0.877 | 0.055 | 0.068 |
| PGM | 0.942 | 0.032 | 0.026 |
| Ak | 0.740 | 0.154 | 0.105 |
| Kidd | 0.763 | 0.218 | 0.020 |
| Duffy | 0.636 | 0.105 | 0.259 |
| Lewis | 0.965 | 0.033 | 0.001 |
| Kell | 0.903 | 0.072 | 0.025 |
| Lutheran | 0.696 | 0.215 | 0.089 |
| P | 0.949 | 0.029 | 0.022 |
| MNS | 0.906 | 0.042 | 0.052 |
| Rh | 0.682 | 0.068 | 0.250 |
| ABO | 0.923 | 0.047 | 0.030 |
| True means | **0.857** | **0.071** | **0.072** |
| L72 Written means | 0.854 | 0.083 | 0.063 |
| L72 Calculated means | 0.861 | 0.076 | 0.076 |

The true, correct apportionment of genetic diversity as recalculated from Lewontin's Table 3, together with the true mean. (APh, Kidd, and P add up to 1.001 and Ak and Lewis to 0.999; the diversity components for all other genes add up to 1; slightly inaccurate summation on rounded errors are sometimes unavoidable.[8]) Lewontin's written mean as indicated in his Table 4, and as extensively quoted, is also provided. However, recalculating just his own Table 4 and ignoring empty cells for among populations and among races diversity components for Ag, Lp, and Xm, gives calculated means of 0.861, 0.076, and 0.076. These averages are unbalanced and add up to 1.013. It is unclear how Lewontin calculated his own written mean values, and it is admittedly curious that he overstated the among populations but within races diversity component at the expense of the among races diversity component.

---

[8] Lewontin adjusted the values of the three diversity components for the relevant 14 genes of Lewontin (1972) Table 4 to add up exactly to 1 (see Winther, 2021).

**Table 1.7** Lewontin Calculation Errors in Percentages

| | Within Pops | Among Pops | Among Races |
|---|---|---|---|
| Hp | 0% | 1.96% | −1.79% |
| Ag | −0.12% | – | – |
| Lp | −0.32% | – | – |
| Xm | 0% | – | – |
| APh | 0.86% | 4.84% | −109.09% |
| 6PGD | −0.23% | 5.17% | −1.49% |
| PGM | 0% | 3.03% | −4.00% |
| Ak | 12.74% | −633.33% | 19.85% |
| Kidd | −2.97% | −3.32% | 58.33% |
| Duffy | 0% | 0% | 0% |
| Lewis | 0.10% | −3.13% | 50.00% |
| Kell | −0.22% | 1.37% | 3.85% |
| Lutheran | −0.29% | −0.47% | 3.26% |
| P | 0% | 0% | 0% |
| MNS | 0.55% | −2.44% | −8.33% |
| Rh | −1.19% | 6.85% | 1.19% |
| ABO | −1.76% | 25.40% | 0.00% |

Deviations of Lewontin's (1972) Table 4, p. 396, from my recalculated Table 1.6, in percentages (a negative number indicates that his number is lower than the actual, true value, and vice versa). Only Ak is absolutely and uniformly off. Most errors are small, and even for some high percentages, the absolute difference is small: e.g., the among races diversity component for Lewis is actually 0.001, whereas Lewontin writes 0.002 (see also Figures 1.4 and 1.5).

---

[1] His Hubby and Lewontin (1965) (properly: 1966) is really two publications: Hubby and Lewontin (1966) and Lewontin and Hubby (1966).

[2] Cf. Beatty (1987). In his influential book, Lewontin drew some provocative contrasts, e.g., between the "pessimistic" classical view and the "optimistic" balance view (Lewontin, 1974, pp. 23–31). Interestingly, Lewontin (1972) also cites Lewontin (1968), which was his first general essay on overarching philosophical and historical aspects of evolution. See his bibliography up to 2001, collated by Instituto Veneto.

[3] For a critical assessment of the influence of his own molecular strategy, see Lewontin (1991). In their retrospective, Charlesworth and Charlesworth (2017) acknowledge Hubby and Lewontin (1966) and Harris (1966) as "the first attempts to quantify genetic variability without any bias towards genes that were already known to be variable" (Charlesworth and Charlesworth, 2017, p. 3). Incidentally, Lewontin (1972) cites Harris (1970), but not Harris (1966) (although Lewontin, 1967a does).

[4] I here abstract away from their extensive and fascinating theoretical discussion of the explanatory, evolutionary forces causing and maintaining allele variation within and among populations (Lewontin and Hubby, 1966, pp. 605–608), which is of less relevance to Lewontin (1972).

[5] Ladislav was of Czech ancestry, born in Vienna, and studied communication theory at MIT, before receiving a Harvard doctorate in applied physics (see: Rechcigl, 2021, loc. 814), and Marie was likely a relation. She is not listed in Rechcigl (2021). See also: http://worldcat.org.ezproxy.uindy.edu/identities/viaf-83694028/ [Accessed 20 May, 2021].

[6] How these tables were themselves calculated is a matter for another time. It is mind-boggling to realize that the first tables of logarithms were produced by the Scottish mathematician John Napier in the early 17th century.

[7] After all, $F$-statistics are premised on $h$; one kind of $F$-statistic, the fixation index (i.e., $F_{ST}$), is a measure also seeking to assess the relative loss or deficit of diversity or heterozygosity in subdivided populations, relative to the total population (see Winther, 2022, Chapters 4 and 5, and references therein).

[8] Regarding "objective techniques" for assessing heterozygosity, Lewontin cites Hubby and Lewontin (1966), Lewontin and Hubby (1966), and Prakash et al. (1969), as well as a mouse study, Selander and Yang (1969), and two human analyses, Lewontin (1967a) and Harris (1970).

[9] He here prefers the term "systems" to "genes" since it was unclear whether some of them involved one gene with multiple alleles or several genes, each with fewer alleles. For simplicity's sake, Lewontin assumes the former. (We know today that this is not true for, e.g., Rh or MNS.)

[10] I verified that averaging the *a* allele for Kidd across Eskimos, "Indians (British Columbians)," Blood, and Blackfoot groups (Mourant's population categories) gives a value very close to Lewontin's stated $a$ = 0.615, in the second row of the Kidd entry of his Table 3. He also states there that there are four populations for Indigenous Americans, consonant with Mourant's Table 39, which, incidentally, is also the only table for Kidd in the three books cited. Notably, Lewontin Table 2 lists "Blackfeet," while Mourant uses "Blackfoot Indians." I suspect Lewontin is using the terms interchangeably, although "All Blackfeet are Blackfoot, but not all Blackfoot are Blackfeet." See: https://indiancountrytoday.com/archive/10-things-you-should-know-about-the-blackfeet-nation [Accessed June 10, 2021].

[11] That these blood groups are still used in antibody panels can be verified here (Xg is added, ABO is absent, otherwise all other eight blood groups are shown): https://commons.wikimedia.org/wiki/File:Serology_interpretation_of_antibody_panel_for_blood_group_antigens.jpg [Accessed May 1, 2021].

[12] For a synoptic, historical presentation of the uses of many of these blood groups, and blood group genes, for medical (e.g., blood transfusions) or forensic (e.g., ancestry or kinship analyses) purposes, starting with the discovery of the ABO blood group system early in the 20th century by Nobel laureate Karl Landsteiner, see Geserick and Wirth (2012).

[13] There is no easy way of extracting allele frequency information from Mourant (1954), Table 40 "Sickle-Cell Trait," pp. 211–240.

[14] Another confusion: In drawing on Giblett's simple tables for recalculating Lp and Xm below, I followed Lewontin's population number, *N*, which forced me to include allele frequencies of (North) Americans with European backgrounds, though they are not listed in Table 1.2. (Although, "U.S. White" is mentioned as one of the four populations for Xm, Lewontin (1972), p. 389. This is the only place in the article where the populations studied for a gene are explicitly listed.) One more: Australian aborigines are represented for only five genes in Table 3, in three cases as a single population, but for Lutheran and MNS, $N = 2$. In the relevant place, Mourant's tables list "New Guinea Natives" for Lutheran (p. 401) and MNS (p. 362), which certainly in the former case is the only other candidate that Lewontin must be thinking of (and allele frequency checking partly verifies this). But Papuans are classified under Oceanians in Table 2.

[15] For instance, Lewontin assumed that the data table allele frequencies of Giblett (1969) and Mourant (1954) were true allele frequency parameters. He did not engage in allele frequency estimation procedures, standard in population genetics. Even so, because the sample sizes of Giblett (1969) and Mourant (1954) were rarely much below 100—typically on the order of several hundred—the data table allele frequencies presented were reliable renditions of the true allele frequencies in the assessed populations (see Winther, 2022).

[16] The Shannon measure of the fixed allele will be 0, since the logarithm of 1 in *any* base is 0. The Shannon measure for all other alleles at the same loci will also be 0, since their gene frequency $p$ is 0. The heterozygosity measure is also 0, since, per eq. 1.1, $h = 0$.

[17] Eqs. 1.1 and 1.2 are *equivalent*; they give the same values $h$ for the same allele frequencies, since all heterozygote and homozygote genotype frequencies must add up to 1.

[18] Inspect Table 3 (Lewontin, 1972) for P, and note that the species level allele frequency is 0.509, meaning that for this biallelic locus, the two allele frequencies are almost equal, and this is reflected in the (maximum) species-averaged diversity value of 1 (cf. Figure 1.2).

[19] This can be verified by his verbal glosses on p. 389 and by recalculating his entire Table 3, which I have done and will summarize below. Note that you can add the words "per locus" to each diversity measure, after the initial "the": e.g., "the per locus diversity of a given population."

[20] For the 17 genes Lewontin assessed, $u = 2$ for 13 genes, $u = 3$ for one gene (ABO blood group), $u = 4$ for two genes (APh, MNS), and $u = 6$ for one gene (the Rhesus factor blood group).

[21] Lewontin observes that counting each population equally overestimates $H_{species}$ "since small populations are given equal weight with large ones," and underestimates $\bar{H}_{pop}$ since "too much weight [is given] to small isolated populations and to less numerous races like [Indigenous Americans] and Australian aborigines, both of which have gene frequencies that differ markedly from the rest of the species" (1972, p. 389). While mathematical consistency obligates Lewontin to state the last phrase, it is unclear how valid this claim is from the Table 3 data presented—inspect haptoglobin, Rh, and ABO and form your own judgment.

[22] The Wahlund effect also holds for more than two alleles at a locus, although with some complexities (e.g., Hartl and Clark, 1989, pp. 288–291; see also Chapter 4 of Nielsen and Slatkin, 2013).

[23] Move the $p_{O2}$ line left or right, any arbitrary amount, even if the diagram "flips" when $p_{O2} < p_{O1}$, and note that $H_{race}$ is always greater than $H_{pop}$, except when the vertical lines coincide. (Alternatively, in your mind's eye, move $p_{O1}$ to the left as you move $p_{O2}$ any arbitrary amount.)

[24] Another comparison of levels could also be captured by another version of this figure or state-space map (Winther, 2020): species-averaged diversity, $H_{species}$ versus the weighted average of every race's $H_{race}$, i.e., $\bar{H}_{race}$. (I allude to this comparison in the "or, alternatively…" parentheses of the main text.) However, I do not here visually represent this comparison, since the diversity values of this pair are necessarily higher than the corresponding values of the represented pair, though not in any fixed way (recall our metaphorical three degrees of freedom, constrained by three inequalities). This is because both $H_{species}$ and $\bar{H}_{race}$ are calculated with more races than the one currently captured by Figure 1.3, races which could have wildly different allele frequencies, thereby significantly increasing $H_{species}$; $\bar{H}_{race}$ too could be quite high, if the internal heterogeneity of the races is high. Incidentally, in this alternative figure, $p_{O1}$ and $p_{O2}$ are replaced, respectively, by $\bar{p}_{r1}$ and $\bar{p}_{r2}$ (the average frequency of the allele in either race), and $\bar{p}_r$ is replaced by $\bar{p}_s$. In such a new rendition, we would move up a level of population structure, and visualize $H_{species} \geq \bar{H}_{race}$.

[25] Formally, this follows from *Jensen's Inequality*, proved by Danish mathematician J. L. W. V. Jensen in 1906, a statement of which is "succinctly written" as eq. 2.6 in Karlin and Taylor (1975), p. 249. The interested reader can verify the geometric overshooting due to convexity with either arithmetical calculations from, e.g., Lewontin (1972), or by confirming the $h_{race} \geq h_{pop}$ inequality, for a biallelic locus, by using $p$ and $(1-p)$ with the simpler *h* measure (eq. 1.1). (Importantly, $H_{race}$ and $H_{species}$ are explicit entropies, and hence are (on) the convex entropy function, while $H_{pop}$ and $\bar{H}_{race}$ are merely averages of, respectively, $H_O$ and $H_{race}$ entropies, and hence fall on the connecting lines.)

[26] That weighting by increasing the number of populations within a race from two to more (or unbalancing the population numbers in different races) does not eliminate the overshooting can also be inferred geometrically from Figure 1.3. Consider the simple case where we have three populations in a race, one of which has allele frequency $p_{O1}$ and two of which have $p_{O2}$. Moving $\bar{p}_r$ horizontally 2/3 rather than 1/2 from $p_{O1}$ towards $p_{O2}$, and doing the same for $H_{pop}$ vertically, it is still the case that $H_{race} > H_{pop}$, though less so. This is so for any weighting of a finite number of populations, i.e., for any extreme difference in numbers of populations within a race with either $p_{O1}$ or $p_{O2}$. (This geometry also holds for the next level of comparison, between $H_{species}$ and $\bar{H}_{race}$; moreover, for this level, consider also that even if some races have 0 populations for some loci, as is the case in Lewontin (1972), those races will not figure in any of the six diversity measures).

[27] Although not quite, for a variety of reasons. Also, Lewontin worries about using analysis of variance (ANOVA) methodology in this context to decompose genetic variance into the three levels, and opts not to employ it (Lewontin, 1972, p. 386; cf. Winther, 2022, Chapters 4 and 5).

[28] Adding eqs. 1.10–1.12 together gives one, since $\frac{H_{species}}{H_{species}} = 1$.

[29] Consult Biddanda et al. (2020) for why this is true, while it is also true that "most variants are rare and geographically localized" (p. 1, ff.).

[30] This distribution is unsurprising, as individuals homozygous for the alternative allele of Table 3 are resistant to the malarial parasite *Plasmodium vivax*, historically common in Africa, and, to a lesser extent, in Southeast Asia, e.g.,

Szpak et al. (2019), pp. 1432–1435. This locus has thus been under selection. A note on nomenclature: while gene names today are typically italicized, also to distinguish them from their protein products (whose names are typically *not* italicized), in this chapter I follow Lewontin's convention of not italicizing gene names. All relevant names refer to genes rather than protein products.

[31] Rosenberg (2011), Figure 3, p. 664 selects three microsatellite loci with roughly analogous levels of allele frequency geographic differentiation, with Duffy corresponding to D12S2070 (bottom row); 6PGD to D6S474 (top row); and P to D10S1425 (middle row).

[32] The overbar on $H_{species}$ in Lewontin's table is unnecessary, as reflected even in his Table 4.

[33] Incidentally, calculating $H_{O,m}$ for every population of 517 total populations must have been an enormous amount of work on the part of Lewontin, and perhaps others. It seems unlikely he could have calculated these, or even rigorously completed all the Table 3 calculations, on a Cambridge, MA–Vermont bus ride (Lewontin, pers. comm., February 17, 2016). I chose not to check for calculation errors, if we can call them that, in most $H_{O,m}$ because often it is unclear which populations he used in each given race, and how he handled them (see "The Samples" section above). Near-endless discussion would have ensued. Rather, it is more powerful to check the internal consistency of Table 3.

[34] Except for Duffy, which involved a single rounding error that did not impact $H_{species}$, and hence did not change the diversity apportionments (see also Figures 1.4 and 1.5, and Table 1.7).

[35] Here I also include Ag and Xm, although he did not make all their diversity components explicit.

[36] See: https://docs.microsoft.com/en-us/office/troubleshoot/excel/floating-point-arithmetic-inaccurate-result [Accessed 1 June, 2021].

[37] Table 34 of Lewontin (1974) corrected some—but only some—of the calculation errors of the 17 genes of Lewontin (1972) (e.g., Lewis), and amended the overarching apportionment of diversity components to 84.9%/7.5%/7.5% (p. 156). This is closer to his own data, but still rounds incorrectly. Moreover, publications focusing on the diversity apportionment of human genetic variation that cite Lewontin (1972) rarely *also* cite Lewontin (1974) (see Winther, 2022).

---

[38] Regarding the last point, given even just the data available to him, Lewontin was obligated to average across loci. Recall that the many studies referred to and tabulated by Giblett (1969) and Mourant (1954) were for specific genes in specific populations, with only some studies looking at multiple populations, and effectively no studies surveying the same individuals for more than one locus—i.e., Lewontin's sources did not include multilocus data. The human genome projects emerging especially in the 1990s of course changed all of that, and introduced oceans of multilocus data.

[39] Although: "six entries, included in error, had to be removed leaving blank numbers" (e.g., #463, p. 265) (Mourant, 1954, p. 239).

[40] See Brown and Armelagos (2001), p. 38, for a useful table comparing eight studies of diversity apportionment subsequent to Lewontin's. Interestingly, and as a small example of the influence of Lewontin (1972), Brown and Armelagos simply report Lewontin's own (incorrect, even by his Table 4) written means, rather than doing any recalculations. See also Table 1 of Barbujani et al. (1997), p. 4518, and Table 1 of Rosenberg et al. (2002), p. 2382. While apportionment always tends to be mostly within populations, depending on the kinds of (autosomal) gene or measure used, especially the among race component can be significant (e.g., 2.8%–14.0%; 7.2%–15.4%; 10%–11.7%, for three studies cited in Brown and Armelagos, 2001, and Jorde et al., 2001 found 10.4%–17.4%; see Winther, 2022, Chapter 6).

Fig. 1.1

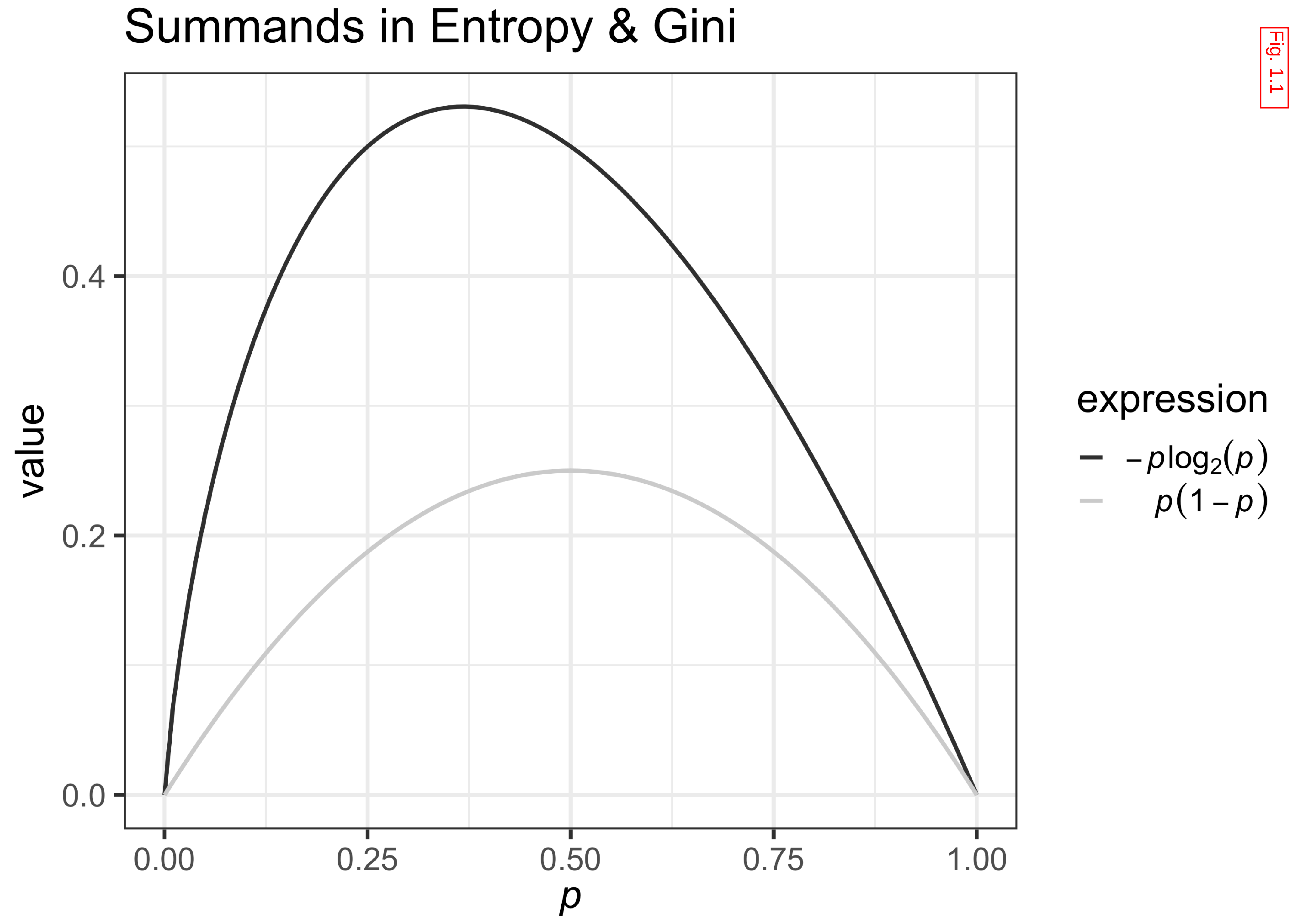

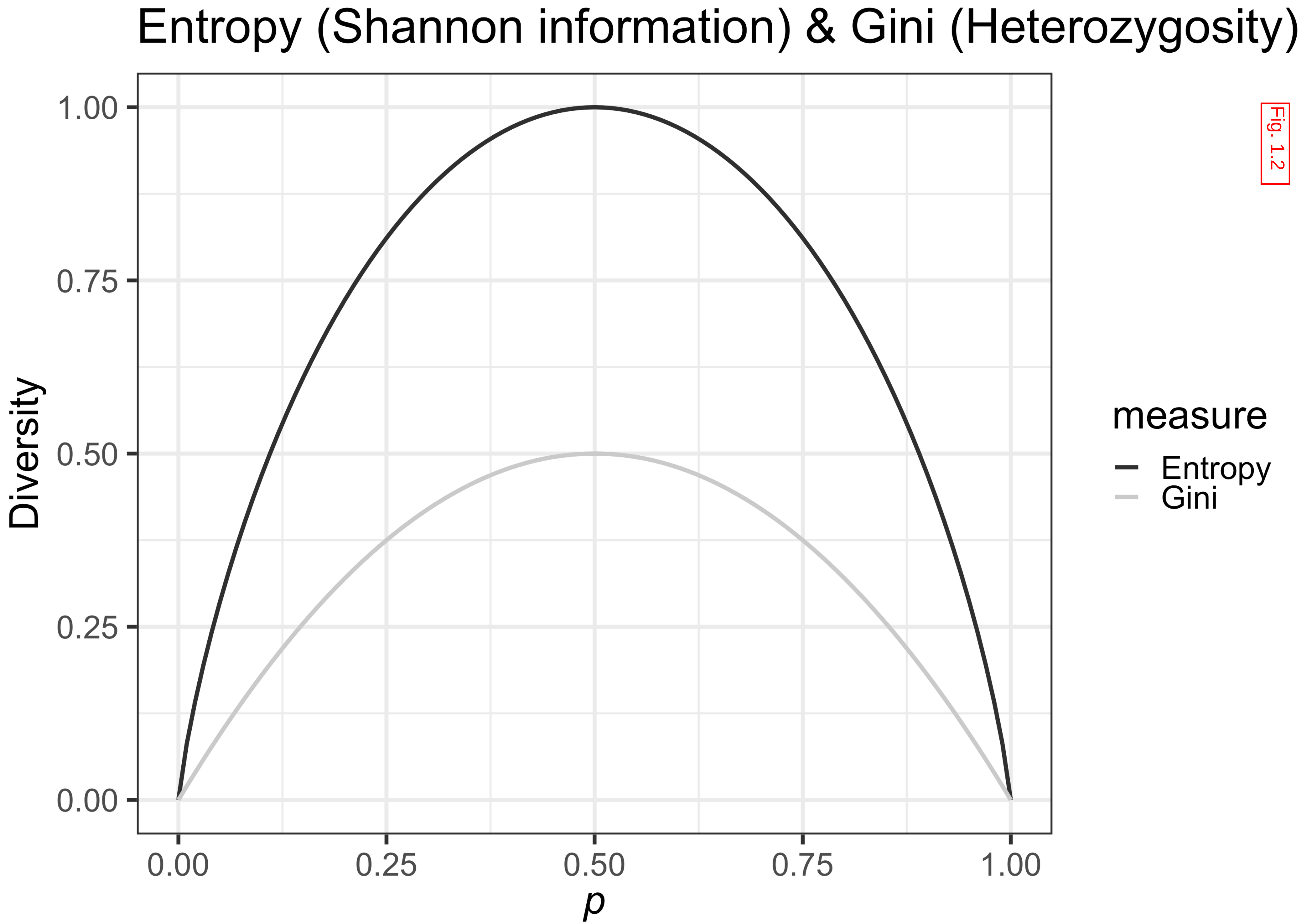

Fig. 1.2

Fig. 1.3

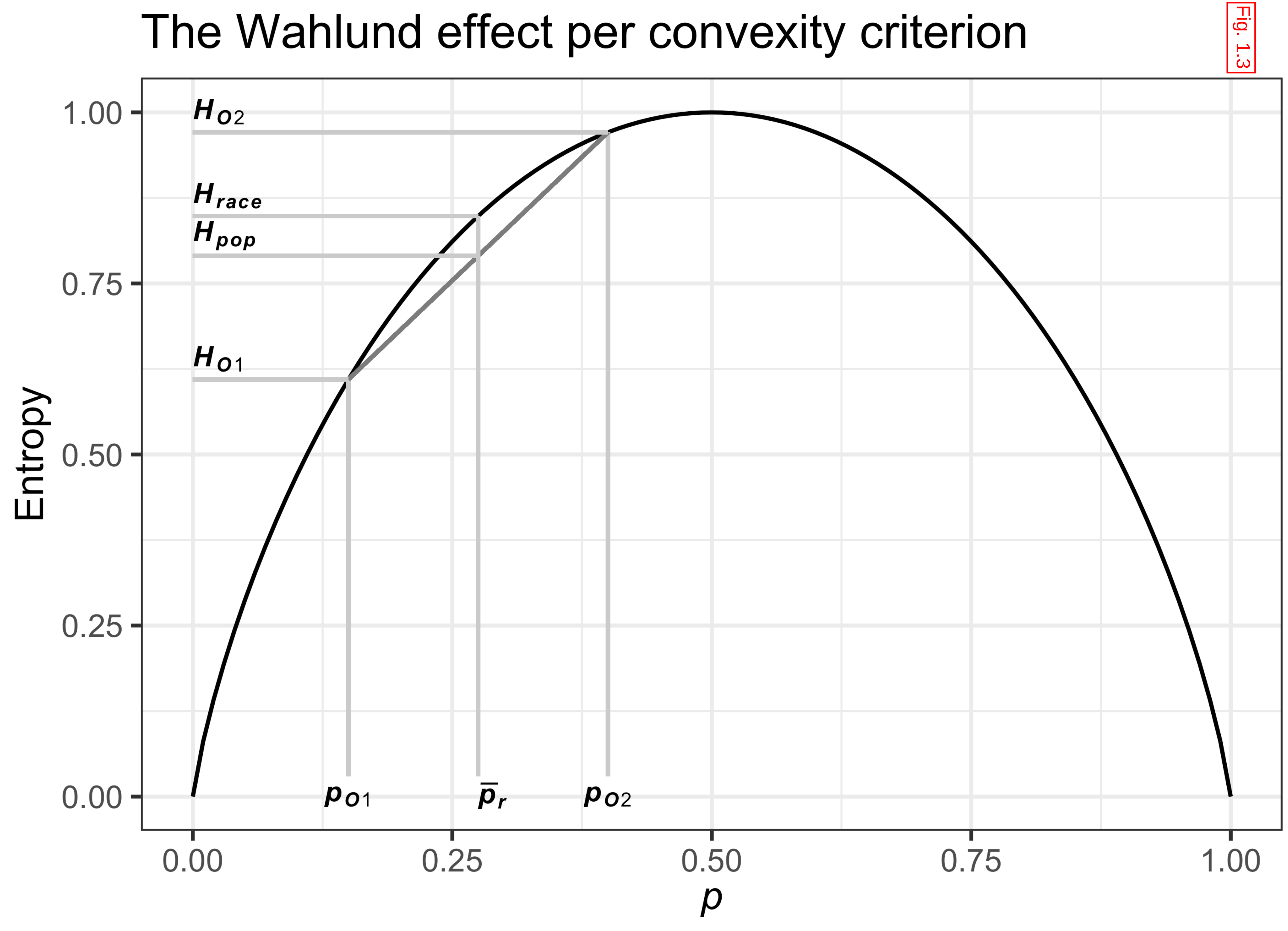

Fig. 1.4

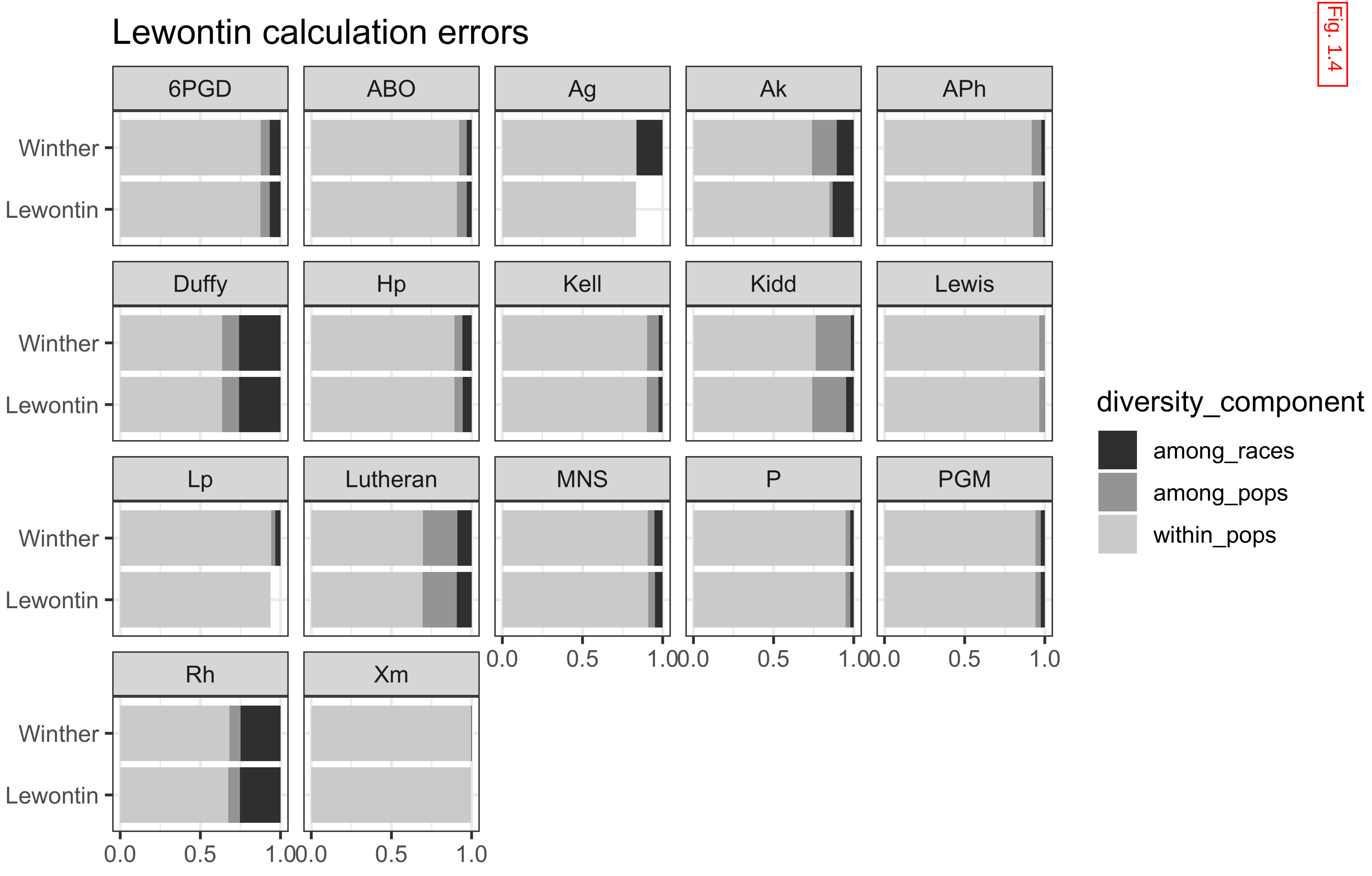

Fig. 1.5

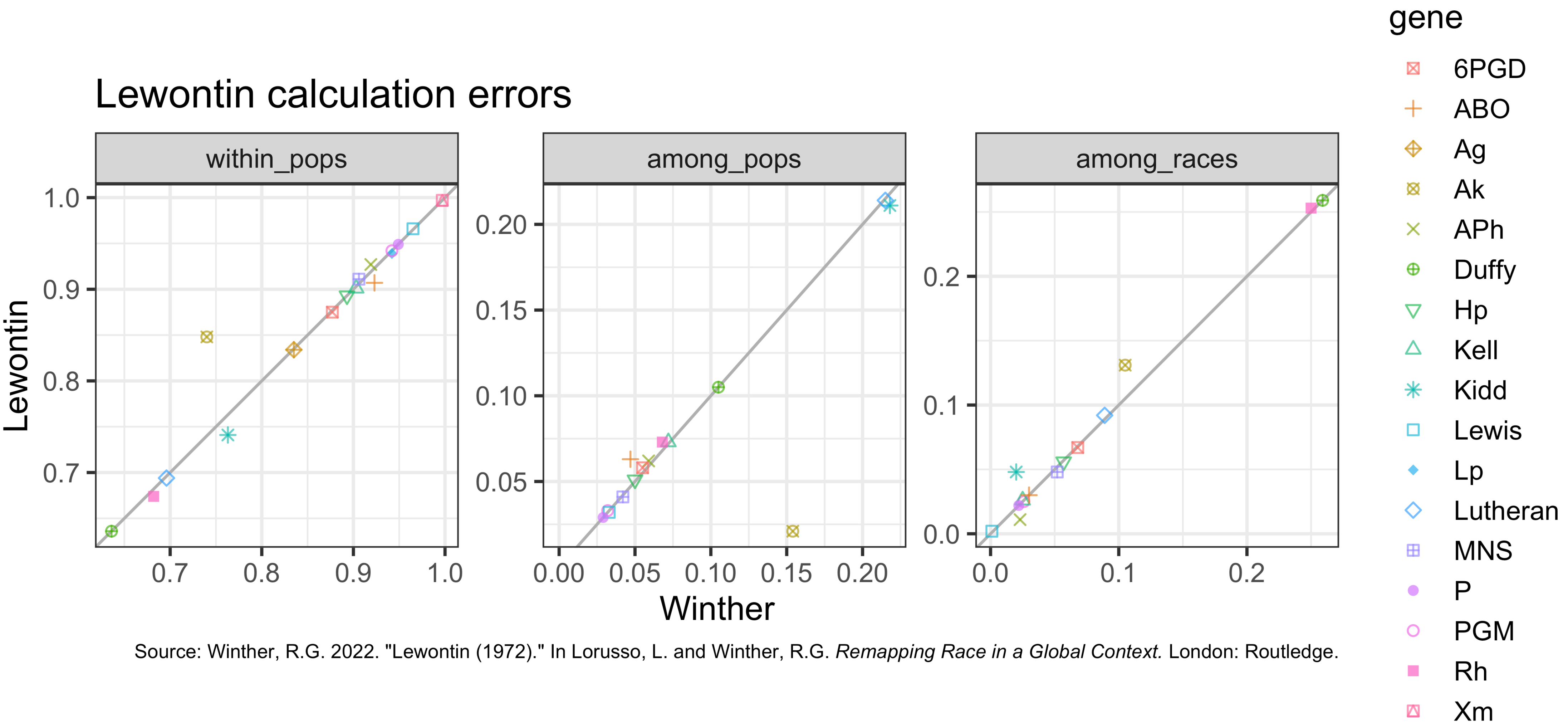

Source: Winther, R.G. 2022. "Lewontin (1972)." In Lorusso, L. and Winther, R.G. *Remapping Race in a Global Context.* London: Routledge.